\documentclass[lettersize,journal]{IEEEtran}
\usepackage{amsmath,amsfonts}
\usepackage{amssymb}
\usepackage{physics,amsmath}
\usepackage{algorithm}
\usepackage{algpseudocode}
\usepackage{array}
\usepackage{bm}
\usepackage{subcaption}
\usepackage{mathtools}
\usepackage{xcolor}
\usepackage{tikz}
\usetikzlibrary{chains}
\usepackage{makecell}
\usepackage{textcomp}
\usepackage{stfloats}
\usepackage{lipsum}  
\usepackage{bm}
\usepackage{url}
\usepackage{verbatim}
\usepackage{graphicx}
\usepackage{cite}
\usepackage{pdfpages}
\usepackage{tikz-3dplot}
\usepackage{relsize}
\usetikzlibrary{positioning, arrows.meta}
\usepackage{xcolor}
\usepackage[hidelinks]{hyperref}
\usepackage{tikz}
\usetikzlibrary{3d}
\usetikzlibrary{arrows.meta} % for more arrow tip kinds

\definecolor{mycustomcolor1}{rgb}{0.6627, 0.1412, 0.1255} 
\definecolor{mycustomcolor2}{rgb}{0.8196, 0.5686, 0.2431}
\definecolor{mycustomcolor3}{rgb}{0.0549, 0.1922, 0.3765}
\definecolor{mycustomcolor4}{rgb}{0.0745, 0.3255, 0.5647}
\definecolor{mycustomcolor5}{rgb}{0.7882, 0.3490, 0.3725} 
\definecolor{mycustomcolor6}{rgb}{0.2157, 0.1843, 0.1843}
\definecolor{mycustomcolor7}{rgb}{0.9843, 0.8078, 0.2275}
\definecolor{mycustomcolor8}{rgb}{0.8784, 0.8157, 0.7216}

\usepackage{courier} % Load the courier package for typewriter font
\usepackage{algorithm} % For algorithms
\usepackage{algpseudocode} % For pseudocode

\algrenewcommand\alglinenumber[1]{\small\ttfamily\textcolor{black}{#1}}
\algrenewcommand\algorithmicrequire{\textbf{\small\ttfamily Input:}}
\algrenewcommand\algorithmicensure{\textbf{\small\ttfamily Output:}}
\algrenewcommand\algorithmiccomment[1]{\hfill\#\ \eqparbox{COMMENT}{\small\ttfamily #1}}

\begin{document}

\newcommand{\orcidiconFeb}{\href{https://orcid.org/0009-0008-2632-1140}{\includegraphics[scale=0.1]{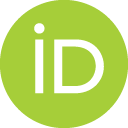}}}

\newcommand{\orcidiconAbk}{\href{https://orcid.org/0009-0006-1187-7782}{\includegraphics[scale=0.1]{figures/orcidID128.png}}}

\newcommand{\orcidiconOba}{\href{https://orcid.org/0000-0003-2523-3858}{\includegraphics[scale=0.1]{figures/orcidID128.png}}}

\title{ Odor Perceptual Shift Keying (OPSK) for Odor-Based Molecular Communication}

\author{Fatih E. Bilgen\orcidiconFeb,~\IEEEmembership{Student Member,~IEEE},
        Ahmet B. Kilic\orcidiconAbk,~\IEEEmembership{Student Member,~IEEE}, 
        and Ozgur B. Akan\orcidiconOba,~\IEEEmembership{Fellow,~IEEE}                
        \thanks{The authors are with the Center for neXt-generation Communications (CXC), Department of Electrical and Electronics Engineering, Koç University, Istanbul, Turkey (e-mail: \{fbilgen20, ahmetkilic20, akan\}@ku.edu.tr).}
        \thanks{Ozgur B. Akan is also with the Internet of Everything (IoE) Group, Electrical Engineering Division, Department of Engineering, University of Cambridge, Cambridge, CB3 0FA, UK (email: oba21@cam.ac.uk).} 
	    \thanks{This work was supported in part by the AXA Research Fund (AXA Chair for Internet of Everything at Ko\c{c} University).}
}% <-this % stops a space


	
\maketitle
\begin{abstract}
Molecular communication (MC) has promising potential and a wide range of applications.  However, odor-based communication which is common in nature, has not been sufficiently examined within the context of MC, yet. In this paper, we introduce a novel approach for implementing odor-based MC systems. We propose a new modulation scheme called Odor Perceptual Shift Keying (OPSK), which encodes information by shifting the perceptual values of odor molecules in pleasantness, intensity and edibility dimensions. We construct a system which transmits OPSK modulated signals between a transmitter and receiver. We conduct analyses on the system parameters to simulate performance metrics such as symbol error rate (SER) and symbol rate (SR). Our analyses indicate that OPSK has a potential for realizing odor-based MC systems. We find that under certain conditions, reliable odor-based MC systems can be implemented using OPSK across a variety of distance ranges from millimeters up to kilometers. Additionally, we introduce adaptive symbol transmission to our system for input symbol sequences featuring symbols that occur with unequal probabilities. We further demonstrate that the proposed algorithm at the transmitter side can achieve extended operation times.

\end{abstract}

\section{Introduction}
\IEEEPARstart{M}{olecular} communication (MC) is the transmission of information using particles such as molecules or lipid vesicles \cite{mcsurvey}. These particles are released into a gaseous or an aqueous medium from transmitter. Propagating particles in the medium reach to the receiver of the system. Arriving particles are detected and decoded to extract the information. Current studies in the field of communication focuses on creating MC systems \cite{akanmcsurvey,mcsurvey,modsurvey}.

In nature, there are variety of MC types. Pheromones are employed for long-range communication, whereas chemical signals are used for intracellular and intercellular communication at micro and nanoscales \cite{mcsurvey}. Utilizing pheromones, or odor molecules, for conveying message is called odor-based communication. We observe this MC type in nature for several species from insects to plants \cite{plantinsect,insect,human}. Mammals using pheromones as information particles is a good example of this phenomenon \cite{mcsurvey,longrangeMC,mammals}. Investigating the details of olfactory communication examples in nature can enable us to develop odor-based communication systems. Studies in this area can be beneficial for biomedical applications, industrial and consumer goods applications, environmental applications, telecommunications, information and communications technology (ICT), and future internet \cite{longrangeMC}. To the best of our knowledge, the existing literature regarding modulation methods in odor-based communication is limited to keying information into concentration levels of odor molecules \cite{odorcsk}. However, to be able to increase the extent of odor communication for diverse applications, new modulation techniques need to be proposed. 

In MC, the five primary types of modulation are concentration-based, type-based, timing-based, spatial, and higher-order modulation techniques \cite{modsurvey}. Concentration-based methods encode information to the concentration of molecules to convey message. However, these methods may be susceptible to inter-symbol-interference (ISI) because of the redundant molecules from previous transmissions \cite{cskmsk}. Keying type of messenger molecules is the idea behind type-based modulation methods. These methods might be more challenging to execute and demands a more complex system design \cite{cskmsk}. 
Information can be expressed by the spatial location of release, particularly for multi-antenna systems in spatial-based modulations \cite{modsurvey}. Timing-based modulations use various time-related features of the carried signal to encode information \cite{modsurvey}. Lastly, hybrid techniques use more than one of the four characteristics of the transmitted signal to convey the message.

Several studies have looked into proposing new methods to tackle issues like ISI and system complexity \cite{shortmod1,shortmod2,shortmod3,shortmod4,shortmod5,shortmod6,shortmod7}. Given the significant differences between MC and traditional communication, realisability and biocompatibility of MC systems are also important considerations \cite{akanmcsurvey}, \cite{kuscutransmittersurvey}. Thus, having alternative options might be beneficial for using MC in a variety of applications. Just like in many situations, nature provides a great source of inspiration. Hereby, understanding the characteristics of odors is critical to come up with new ideas for modulation techniques. 

Studies in olfaction began as early as the 16th century. In one of the first classification schemes, 7 descriptors were used to classify odors \cite{olfactionhistory}. While similar schemes continued to be proposed until the 20th century, a noteworthy classification system was proposed, where a three-dimensional space was created to describe every odor using 6 primary odors \cite{olfactionhistory}. In the mid-20th century, it was suggested in \cite{dyson} that the differences between odor molecules can be explained by their molecular vibration frequencies. Even though the notion made a lot of sense, there was no strong evidence on how the nose might really pick up on the vibrations of odor molecules. Later on, studies on olfaction focused more on the physical attributes of odor molecules \cite{olfactionhistory}. According to the stereochemical theory of odor, perception of an odor is influenced by its structure, namely its size and shape \cite{stereochemical}. During the 1960s, it was proposed that olfactory sensing relies on chemosensors, which combine matching depressed sites, hydrogen bonding, dipoles, and van der Waals-London dispersion forces \cite{dravnieks}. Hence, in \cite{dravnieks} a physicochemical basis of odors was created. Furthermore, by creating a numeric dataset on descriptors of odors using the data from the experiments, a basis for further studies on odor perception was laid down \cite{pleasantnesslinear,dravnieksatlas}. In addition to this, similar datasets have also been proposed \cite{pleasantnesslinear}. In 1968, an odor set consisting of 51 compounds was created \cite{harper}. Additionally, other datasets were also created \cite{moncrieff,wrightdata,amoore1967correlations}. By applying statistical methods on these datasets, valuable insights on perception of odors have been gained \cite{pleasant,zarzopleasantnessedibility,pleasantnesslinear}. In \cite{pleasant}, by applying PCA analysis on physicochemical and perceptual space of odors, it has been observed that pleasantness is the primary dimension of the perceptual space of odors and there is a correlation between perceptual and physicochemical space of odors. In this paper, edibility was also thought as another dimension of odor perceptual space. Similarly, in \cite{intensitypercept,intensityperceptual,intensityperceptual2}, it was understood that intensity can be taken as another dimension of odor perceptual space. Current studies in olfaction seek for mappings from physicochemical space to perceptual space \cite{zarzopleasantnessedibility,pleasant,intensity}. In literature, there exists linear regression formulas that connects two spaces to each other \cite{pleasant,pleasantnesslinear}. However, to the best of our knowledge, studies on odor as a communication system is lacking and there is limited study regarding modulation methods on odor communication.

In this paper, we propose a new modulation scheme for odor-based communication called Odor Perceptual Shift Keying (OPSK).  This paper is structured as follows: Section II explains the OPSK technique. Additionally, several system parameters are defined in Section II. In Section III, we introduce our proposed algorithm for implementing adaptive symbol transmission. The effects of system parameters on performance metrics are analyzed in Section IV. Finally, conclusions are drawn in Section V.

\section{Odor Perceptual Shift Keying (OPSK)}

OPSK encodes information onto the perceptual dimensions of odors. These perceptual dimensions are pleasantness, intensity, and edibility. In perceptual terms, pleasantness refers to the degree to which an odor is liked when smelled. Intensity, on the other hand, relates to the perception of the odor's sharpness upon smelling. Meanwhile, edibility measures the extent to which an odor evokes the desire to eat the source of the odor. 

In \cite{intensity,enosepleasantness}, it is shown that pleasantness and intensity values of odors can be calculated using e-noses. Consequently, we assume each odor possesses values for pleasantness, intensity, and edibility, which can be mapped on a scale from 0 to 100. On this scale, odors inducing the highest degree of pleasantness when smelled by individuals receive the highest values, whereas odors inducing the least pleasant sensations obtain the lowest values. This method of evaluation applies equally to the intensity and edibility perceptual dimensions.

The perceptual values that odors possess in each perceptual dimension enable the creation of different classes of odors. The use of odors from different classes allows for the representation of various combinations of bit sequences. This is the main idea behind the OPSK. In OPSK, digital information is encoded onto a modulated signal by shifting the perceptual values of odors across the dimensions of pleasantness, intensity, and edibility. 

In this section, we provide a comprehensive overview of the point-to-point odor-based MC system, which employs OPSK for modulation. The process at the transmitter involves a step-by-step transformation of a digital signal into an OPSK-modulated signal. Subsequently, the propagation of this modulated signal through the communication channel is examined, including its interaction with channel noise. Finally, the details of the procedures for demodulating and decoding the received signal are given.

\subsection{OPSK Modulation}\label{section:Transmitter} 
The fundamental function of the transmitter is to transform the input digital signal into OPSK-modulated signal. The digital signal, consisting of a sequence of bits, is divided into groups, each containing $K$ bits. In each symbol interval, a group of $K$ bits associated with a specific odor is transmitted. Of these $K$ bits, $n_p$ bits are represented by the odor's pleasantness value, while the remaining $n_i$ and $n_e$ bits correspond to its intensity and edibility values, respectively. The subsequent subsections will explore how OPSK is executed at the transmitter. Initially, we describe the method for selecting the odors of system. The following part details the signal encoding process, and the final part explains the generation of OPSK-modulated signals to be released from the transmitter.

\subsubsection{Transmitter Design}{\label{Transmitter:Initialization}}
We assumed that perceptual dimensions, pleasantness, intensity, and edibility, of an odor can be assessed within the scale of $0$ to $100$. Therefore, we use the following vector notation to denote each odor : $(p_O$, $i_O$, $e_O)$ where $p_O$, $i_O$, $e_O$ $\in [0, 100]$. This vector can be called the perceptual value vector of an odor. 

In OPSK, by leveraging the distinct perceptual value vectors associated with various odors, it is possible to classify the odors. Dividing the assessment scale of a perceptual dimension into intervals create different classes within this perceptual dimension. The combination of different classes from every perceptual dimension creates distinct regions within the perceptual vector space. Hereby, each odor can be identified with the region it belongs in the perceptual vector space.

To be able construct different classes in a perceptual dimension, threshold values that define the intervals within the assessment scale need to be established. Let $p_T$, $i_T$, and $e_T$ be the sets that contain threshold values of each dimension. Then these sets can be created as
\begin{equation}
p_{T} = \left\{ \frac{100}{2^{n_p}}, 2 \cdot \frac{100}{2^{n_p}}, \ldots, (2^{n_p} - 1) \cdot \frac{100}{2^{n_p}} \right\},
\label{eq:pt}
\end{equation}
\begin{equation}
i_{T} = \left\{ \frac{100}{2^{n_i}}, 2 \cdot \frac{100}{2^{n_i}}, \ldots, (2^{n_i} - 1) \cdot \frac{100}{2^{n_i}} \right\},
\label{eq:it}
\end{equation}
\begin{equation}
e_{T} = \left\{ \frac{100}{2^{n_e}}, 2 \cdot \frac{100}{2^{n_e}}, \ldots, (2^{n_e} - 1) \cdot \frac{100}{2^{n_e}} \right\},
\label{eq:et}
\end{equation}
where $n_p$, $n_i$, and $n_e$ represent the number of bits to be transmitted with pleasantness, intensity, and edibility dimensions, respectively.

\begin{figure}[t]
    \centering
    \begin{tikzpicture}
        % Define block style
        \tikzstyle{block} = [draw, rectangle, minimum width=1cm, minimum height=1cm]

        % Block 1
        \node[block] at (0,0) (block1) {};
        \foreach \i in {1,...,50} {
            \fill[mycustomcolor1] (block1.center)++(rand*0.3,rand*0.3) circle (0.08);
        }
        \node[block,label={[align=center, , font=\fontsize{8}{10}\selectfont]below:$(10, 15, 5)$ \\ $O_{000}$}] at (block1) {};
         
        % Block 2
        \node[block, right=1cm of block1] (block2) {};
        \foreach \i in {1,...,50} {
            \fill[mycustomcolor2] (block2.center)++(rand*0.3,rand*0.3) circle (0.08);
        }
        \node[block,label={[align=center, , font=\fontsize{8}{10}\selectfont]below:$(5, 30, 70)$ \\ $O_{001}$}] at (block2) {};

        % Block 3
        \node[block, right=1cm of block2] (block3) {};
        \foreach \i in {1,...,50} {
            \fill[mycustomcolor3] (block3.center)++(rand*0.3,rand*0.3) circle (0.08);
        }
        \node[block,label={[align=center, , font=\fontsize{8}{10}\selectfont]below:$(40, 60, 10)$ \\ $O_{010}$}] at (block3) {};

        % Block 4
        \node[block, right=1cm of block3] (block4) {};
        \foreach \i in {1,...,50} {
            \fill[mycustomcolor4] (block4.center)++(rand*0.3,rand*0.3) circle (0.08);
        }
        \node[block,label={[align=center, , font=\fontsize{8}{10}\selectfont]below:$(1, 95, 95)$ \\ $O_{011}$}] at (block4) {};

        % Block 5
        \node[block, below=1 cm of block1] (block5) {};
        \foreach \i in {1,...,50} {
            \fill[mycustomcolor5] (block5.center)++(rand*0.3,rand*0.3) circle (0.08);
        }
        \node[block,label={[align=center, , font=\fontsize{8}{10}\selectfont]below:$(90, 10, 35)$ \\ $O_{100}$}] at (block5) {};

        % Block 6
        \node[block, right=1cm of block5] (block6) {};
        \foreach \i in {1,...,50} {
            \fill[mycustomcolor6] (block6.center)++(rand*0.3,rand*0.3) circle (0.08);
        }
        \node[block,label={[align=center, , font=\fontsize{8}{10}\selectfont]below:$(70, 25, 95)$ \\ $O_{101}$}] at (block6) {};

        % Block 7
        \node[block, right=1cm of block6] (block7) {};
        \foreach \i in {1,...,50} {
            \fill[mycustomcolor7] (block7.center)++(rand*0.3,rand*0.3) circle (0.08);
        }
        \node[block,label={[align=center, , font=\fontsize{8}{10}\selectfont]below:$(65, 85, 5)$ \\ $O_{110}$}] at (block7) {};

        % Block 8
        \node[block, right=1cm of block7] (block8) {};
        \foreach \i in {1,...,50} {
            \fill[mycustomcolor8] (block8.center)++(rand*0.3,rand*0.3) circle (0.08);
        }
        \node[block,label={[align=center, , font=\fontsize{8}{10}\selectfont]below:$(90, 85, 95)$ \\ $O_{111}$}] at (block8) {};
    \end{tikzpicture}
    \caption{An odor bank example for $n_p=1$, $n_i=1$, $n_e=1$. Each odor is represented by its perceptual value vector $(p_O$, $i_O$ ,$e_O)$ and with the corresponding odor class according to (\ref{eq:perceptual_code}). Different colors and shapes are only for visual purposes.}
    \label{fig:odorbank}
\end{figure}
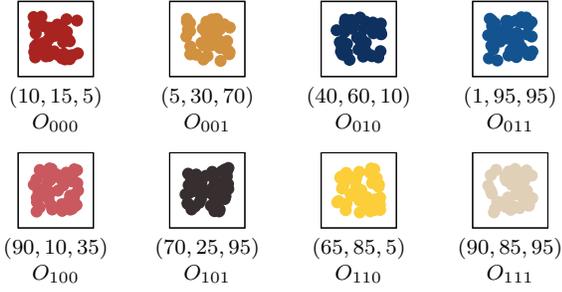

Sets in (\ref{eq:pt}-\ref{eq:et}), create $2^{n_p}$, $2^{n_i}$, and $2^{n_e}$ classes in each perceptual dimension, with each class defined by intervals of equal length. All possible combinations of different classes create $ 2^{n_p}\cdot2^{n_i}\cdot2^{n_e} $ regions within the perceptual vector space. To distinguish these regions from each other, a 3-digit class code $O_{pie}$ can be created. The first digit, $p$, of this code represent the pleasantness class whereas the second digit, $i$, represent the intensity class, and the third digit, $e$, represent the edibility class. These digits, $p$, $i$, and $e$, can take integer values in the range of $[0, 2^{n_p} - 1]$, $[0, 2^{n_i} - 1]$, and $[0, 2^{n_e} - 1]$, respectively. Then for each odor class, we need to match an odor whose perceptual value vector fall within the region of corresponding class according to
\begin{equation}
    O_{pie} = \left\{
        \begin{aligned}
            p &= \left\lfloor\frac{2^{n_p}}{100} \cdot p_O\right\rfloor,\hspace{0.5cm}  0 \leq p_O < 100\\
            i &= \left\lfloor\frac{2^{n_i}}{100} \cdot \hspace{0.5mm}i_O\right\rfloor,\hspace{0.5cm}  0 \leq\hspace{0.5mm}i_O < 100 \\
            e &= \left\lfloor\frac{2^{n_e}}{100} \cdot e_O\right\rfloor,\hspace{0.5cm}   0 \leq e_O < 100 
        \end{aligned}\right.,
    \label{eq:perceptual_code}
\end{equation}
where $(p_O, i_O, e_O)$ is the perceptual value vector of odor and $n_p$, $n_i$, and $n_e$ are the number of bits for each dimension. Let the name of the transmitter part where different odors held at different capsules be odor bank. Fig. \ref{fig:odorbank} illustrates an odor bank for $n_p = 1$, $n_i = 1$, $n_e = 1$.

An odor can comprise either numerous types of molecules or a singular type of molecule \cite{olfactionbase}. In OPSK, we assume that odors matched with the odor class are single type molecular entities. Upon identifying an odor for each class, we proceed to store them within the capsules located at the transmitter. The specific amount of mass to capsulate for each odor is determined and allocated accordingly. This completes the initialization of transmitter. 

\subsubsection{Encoding}\label{Transmitter:Encoding}
In traditional communication systems, symbols are generated from various bit sequence combinations. The total number of symbols needed changes according to the number of bits each symbol represents. A direct one-to-one correspondence exists between the possible bit sequences and the symbols that represent them. Similarly, in OPSK, a one-to-one relationship is established between symbols and the odor classes in the odor bank, with each odor class representing a unique symbol. These symbols are derived from the information conveyed in bit stream form. The bit stream is divided into groups of $K$ bits. Each bit group corresponds to an odor selected from the odor bank. The odors within the odor bank are categorized by class codes which are assigned according to (\ref{eq:perceptual_code}). To select a unique odor for each bit group, we further split the \(K\) bits into smaller segments. The decimal equivalents of bit segments are used to decide the class of odor. In this case, the initial $n_p$ bits of the $K$ bits are used to determine the pleasantness class, the subsequent $n_i$ bits are used to determine the intensity class, and the last $n_e$ bits are used to determine the edibility class. This is achieved by converting the bit segments into their decimal equivalents. Once the classes for each dimension is identified, the bit group is matched with the corresponding odor class in the odor bank. The odor representing the odor class is then released into the channel.

\begin{figure}[t]
    \centering
    \begin{subfigure}{0.52\linewidth}
        \includegraphics[height=4cm]{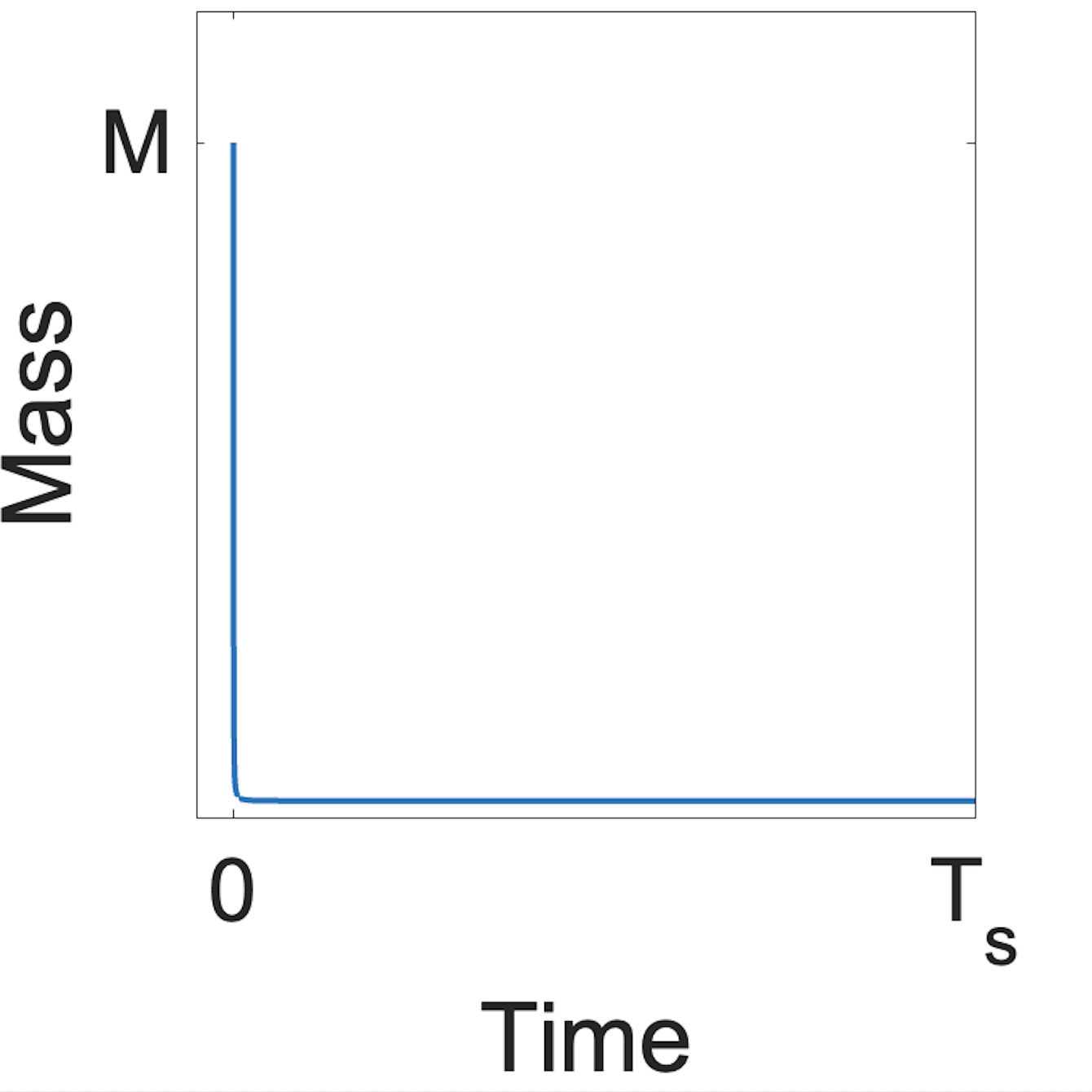}
        \caption{$s_n(t)$}
        \label{fig:waveformSingle}
    \end{subfigure}
    \begin{subfigure}{0.42\linewidth}
        \includegraphics[height=4cm]{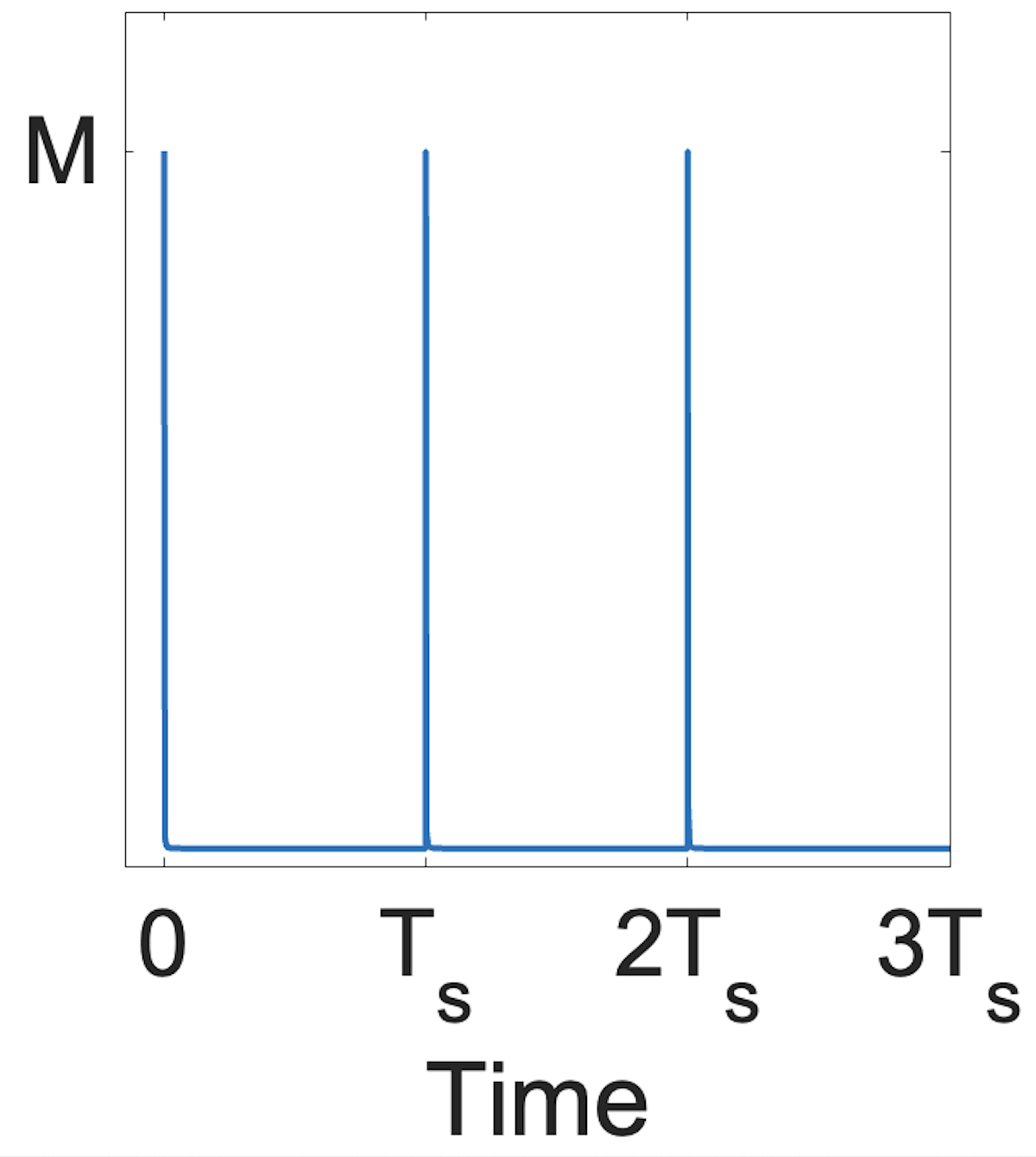}
        \caption{$s(t)$}
        \label{fig:waveformMultiple}
    \end{subfigure}  
    \caption{(a) The waveform of OPSK-modulated signals. (b) The output of transmitter when a sequence of symbols is transmitted.}
    \label{fig:waveform} 
\end{figure}

\subsubsection{Waveform\label{Transmitter:Waveform}}
The release of odor molecules from the transmitter is assumed to be an instant release of total mass $M$ from odor belonging to the class $O_{{pie}}$. $M$ is assumed to be the same for each odor release. Assuming transmitter is located at the coordinates $(0, 0, 0)$, the total mass of odor molecules at any arbitrary time $t$ and at location $(0, 0, 0)$ can be derived by taking the volume integral of concentration function over the smallest volume that includes the transmitter. Let $s_n(t)$, as shown in Fig. \ref{fig:waveform}(a), be the signal for the $n^{th}$ symbol. Then it can be expressed as
\begin{equation}
s_n(t) = \int_{-x_t}^{x_t} \int_{-y_t}^{y_t} \int_{-z_t}^{z_t}C_{n}(0, 0, 0, t) \,dx\,dy\,dz, 
\end{equation}
where $-x_t$, $x_t$, $-y_t$, $y_t$,  $-z_t$, and $z_t$ are the corners of the smallest rectangular volume that includes transmitter and $C_n(x,y,z,t)$ denotes the concentration function of odor molecules released for the $n^{th}$ symbol. Then the output signal of transmitter for sequence of symbols, as shown in Fig. \ref{fig:waveform}(b), can be stated as
\begin{equation}
s(t) = \sum_{n=1}^L s_n(t).
\end{equation}

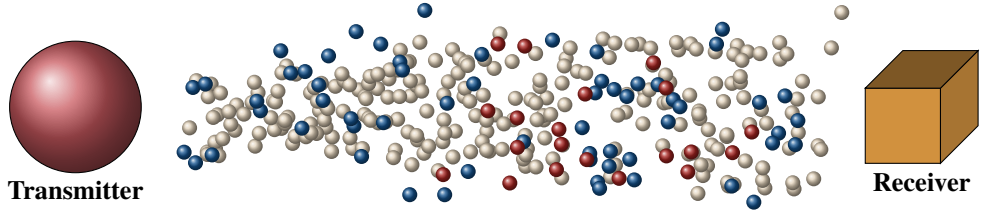
\begin{figure*}[t]
    \centering
    \begin{tikzpicture}[]
        % Your existing TikZ code here
        \node[draw, circle, minimum size = 1.75cm, font=\sffamily,shading=ball, ball color=mycustomcolor5] (transmitter) at (-2.5, 0, 0) {\textbf{}};

        \node at (-2.5,-1.1,0) {\textbf{Transmitter}};
        
         % Node position
        \coordinate (receiver) at (8,-0.75,0);

        % Drawing a cube
        \begin{scope}[shift={(receiver)}]
            \draw[fill=mycustomcolor2] (0,0) rectangle ++(1,1);
            \draw[fill=mycustomcolor2!80!black] (1,1) -- ++(0.5,0.5) -- ++(0,-1) -- ++(-0.5,-0.5) -- cycle;
            \draw[fill=mycustomcolor2!60!black] (1,1) -- ++(0.5,0.5) -- ++(-1,0) -- ++(-0.5,-0.5) -- cycle;
        \end{scope}
    
        % Label
        \node at (8.75,-1,0) {\textbf{Receiver}};
       
        % Draw odor particles as shaded spheres        
        \foreach \i in {1,...,80} {
            \pgfmathsetmacro{\xcoord}{ 0.5 + rand*1.4}
            \pgfmathsetmacro{\ycoord}{ 0   + rand*0.6}
            \pgfmathsetmacro{\zcoord}{ 0   + rand*0.4}
            \shade[ball color=mycustomcolor8] (\xcoord, \ycoord, \zcoord)   circle (0.1);
        } 
        \foreach \i in {1,...,80} {
            \pgfmathsetmacro{\xcoord}{ 3   + rand*1.4}
            \pgfmathsetmacro{\ycoord}{ 0   + rand*0.8}
            \pgfmathsetmacro{\zcoord}{ 0   + rand*0.6}
            \shade[ball color=mycustomcolor8] (\xcoord, \ycoord, \zcoord)   circle (0.1);
        }  
        \foreach \i in {1,...,80} {
            \pgfmathsetmacro{\xcoord}{ 6   + rand*1.4}
            \pgfmathsetmacro{\ycoord}{ 0   + rand*1}
            \pgfmathsetmacro{\zcoord}{ 0   + rand*0.8}
            \shade[ball color=mycustomcolor8] (\xcoord, \ycoord, \zcoord)   circle (0.1);
        }
        \foreach \i in {1,...,60} {
            \pgfmathsetmacro{\xcoord}{ 3   + rand*4}
            \pgfmathsetmacro{\ycoord}{ 0   + rand*1}
            \pgfmathsetmacro{\zcoord}{ 0   + rand*1}
            \shade[ball color=mycustomcolor4] (\xcoord, \ycoord, \zcoord)   circle (0.1);
        }
        \foreach \i in {1,...,20} {
            \pgfmathsetmacro{\xcoord}{ 4.5 + rand*2}
            \pgfmathsetmacro{\ycoord}{ 0   + rand*1}
            \pgfmathsetmacro{\zcoord}{ 0   + rand*1}
            \shade[ball color=mycustomcolor1] (\xcoord, \ycoord, \zcoord)   circle (0.1);
        }
    \end{tikzpicture}
    \caption{Propagation of odor molecules in the channel.}
    \label{fig:odor_diffusion_aligned}
\end{figure*}

\subsection{OPSK Channel}\label{section:Channel}
In nature, odor communication occurs in an open-air channel. In this channel, the movement of molecules has two sources: first, the diffusion of particles, and the second, additional flows such as wind that increase the extent of information over longer distances \cite{tuncparam}. Therefore, for our system, we also have chosen an open-air channel model with an additional flow. 

\subsubsection{Noise in the Channel}\label{Channel:Noise}
Since odor molecules propagate in an open-air channel, external factors that affect the motion of molecules can not be neglected. Any kind of intervention to molecules' movement introduces another source of motion that can be thought as the noise in the channel. This noise can be included into system as deviations from the magnitude of flow rate. Let's revise the second source of motion in the system as average flow rate. Then, average flow rate can be modelled as a Gaussian random variable. Hence, the average flow rate vector of channel at each symbol interval can be expressed as
\begin{equation}
    \boldsymbol{\vec{U_{{avg}}}} = ( N(v_x, \sigma_x^2), N(v_y,    \sigma_y^2) , N(v_z, \sigma_z^2) ),
    \label{eq:averageFlowRate}
\end{equation}
where $v_x$, $v_y$, $v_z$ are the created flow rates on $x$, $y$, $z$ directions, and $\sigma_x$, $\sigma_y$, $\sigma_z$ are the standard deviations from the flow rates on $x$, $y$, $z$ directions. Since external factors change with time, we assume different average flow rate for each symbol interval. Thus, the average flow rate for each symbol interval can be given as
\begin{equation}
\boldsymbol{\vec{u}} =  sample(\boldsymbol{\vec{U_{{avg}}}}), 
\label{eq:samplingAverageFlowRate}
\end{equation}
where $\boldsymbol{\vec{U_{{avg}}}}$ defined in (\ref{eq:averageFlowRate}).
We defined the ratio of created flow rate to standard deviation from this value as our noise parameter. We named this parameter as flow rate to noise ratio which is denoted by $FNR$. Due to possible differences in flow rate and noise on different directions, $FNR$ can be defined in each direction separately. Therefore, on the x direction it can be expressed as
\begin{equation}
    FNR_x = \frac{v_x}{\sigma_x},
    \label{eq:FNR}
\end{equation}
where $v_x$ represents the flow rate and $\sigma_x$ represents the standard deviation from flow rate $v_x$. $FNR_y$ and $FNR_z$ can also be expressed similarly. 

\subsubsection{Propagation in the Channel}{\label{Channel:Propagation}}
Concentration of molecules in relation to diffusion and advection is articulated in \cite{advectioneq} as
\begin{equation}
    \frac{\partial C}{\partial t} = D \cdot \nabla^2 C - \nabla \cdot (\boldsymbol{\vec{u}} C),
    \label{eq:pde}
\end{equation}
where $C$ is the concentration function in space-time, $D$ is the diffusion coefficient of odor molecules for the given medium, and $\boldsymbol{\vec{u}} = (u_x, u_y, u_z)$ is the average flow rate vector in (\ref{eq:samplingAverageFlowRate}). 

An instant release of odor molecules of mass $M_{n}$ for the $n^{th}$ symbol, at $t = 0$, from transmitter located at $(0, 0, 0)$, can be taken as the initial condition to characterize the solution of (\ref{eq:pde}) as
\begin{equation}
    C_n(x, y, z, t) = \frac{M_{n}}{\left(4\pi t\right)^{3/2} \sqrt{D_x D_y D_z}} \cdot e^{-\sum_{k \in \{x, y, z\}} \frac{(k - u_{\text{k}} \cdot t)^2}{4D_{k}t}},
    \label{eq:concentration}
\end{equation}
where $C_n$ represents the concentration function of odor molecules that are released from transmitter for $n^{th}$ symbol, $D_x, D_y,$ and $D_z$ are diffusion coefficients on $x$, $y$, and $z$ directions,  $u_x, u_y, u_z$ are average flow rates on $x$, $y$, and $z$ directions. The unit of $C_n$ is kg/m$^3$. Fig. \ref{fig:odor_diffusion_aligned}, demonstrates the propagation of odor molecules in the channel. 

\subsection{OPSK Receiver}\label{section:Receiver} 
Many types of receivers have been developed for MC, catering to various applications within the field \cite{kuscutransmittersurvey}. Inspired by the existing receiver types in the literature, we assumed a passive-absorber receiver type.  At absorption time in a symbol interval, a passive-absorber receiver can absorb the molecules within a defined volume. We assume that our receiver is located at $(x_{r}, y_{r}, z_{r})$ and it absorbs molecules from the cube-shaped volume that has edge length $l$.  

Upon completion of absorption, the receiver holds a sample of the molecules present in the channel. During each symbol interval, it is expected that the molecules contributing most to the absorbed sample correspond to those released for the most recent symbol transmission. However, external factors introducing noise into the channel can result in the presence of leftover molecule types from previous transmissions. Consequently, a filtering process is necessary to remove these irrelevant molecules before analyzing the absorbed sample. This filtration involves identifying the masses of different molecule types and selecting the one with the greatest mass from the absorbed sample. Therefore, choosing the optimal absorption timing, when the concentration of molecules released for the most recent symbol interval is at its peak within the absorption volume, is crucial.

The analysis of the absorbed symbol after filtering includes demodulation and decoding stages. During demodulation, the goal is to calculate the perceptual value vector of the odor, while the decoding stage aims to extract the bit sequence from the odor class of the calculated perceptual value vector according to (\ref{eq:perceptual_code}). Once the bit sequence is successfully extracted, symbol transmission for that interval is considered complete. The subsequent subsections provide a detailed examination of each stage.

\subsubsection{Absorption}\label{Receiver:Absorption}
To find the total absorbed mass at a specific time we need to take the volume integral of concentration function within the borders of absorption volume.  Let  $T_n$  represent the total amount of time that elapsed between the beginning of the first symbol transmission and the absorption time in the $n^{th}$ symbol interval. Then $T_n$ can be expressed as
\begin{equation}
    T_n = (n-1) \cdot T_s + T_a 
    \label{eq:absorptionTimeN},
\end{equation}
where the time between each symbol transmission is $T_s$ and absorption time in each symbol interval is $T_a$. Hence, total absorbed mass in the $n^{th}$ symbol can be calculated with
\begin{equation}
    \begin{aligned}
    M_{a_n} = \iiint_{\mathlarger{\nu}}\sum_{k=1}^n C_{n+1-k}(x, y, z, T_k) \,dx\,dy\,dz
    \end{aligned}
    \label{eq:absorptionIntegral},
\end{equation}
where $M_{a_n}$ is the total absorbed mass in the $n^{th}$ symbol interval and $T_k$ represent the total amount of time that elapsed between the beginning of the first symbol transmission and the absorption time in the $k^{th}$ symbol interval. It can also be expressed as
\begin{equation}
    T_k = (k-1) \cdot T_s + T_a 
    \label{eq:absorptionTimeK}.
\end{equation}
Since integral and summation are linear operations, we can change their order. In addition to this, replacing $C_n$ in (\ref{eq:absorptionIntegral}) with the expression in (\ref{eq:concentration}) and organizing the terms give us 
\begin{equation}
    \begin{aligned}
    M_{a_n} = \sum_{k=1}^n &\frac{M_{n+1-k}}{\left(4\pi T_k\right)^{3/2} \sqrt{D_x D_y D_z}}\\&\iiint_{\mathlarger{\nu}} e^{-\sum_{i \in \{x, y, z\}} \frac{(i - u_{\text{k}} \cdot T_k)^2}{4D_{i}T_k}}\,dx\,dy\,dz.
    \end{aligned}
    \label{eq:mn_first}
\end{equation}
Since each term in the exponent of $e$ is only dependent on either $x$, $y$ or $z$, integrals in each dimension can be calculated independently and (\ref{eq:mn_first}) can be rewritten as 
\begin{equation}
    \begin{aligned}
    M_{a_n} = \sum_{k=1}^n \frac{M_{n+1-k}}{\left(4\pi T_k\right)^{3/2} \sqrt{D_x D_y D_z}}\cdot I_x \cdot I_y \cdot I_z,
    \end{aligned}
\end{equation}
where $I_x$ is
\begin{equation}
I_x = \int_{x_r - \frac{l}{2}}^{x_r + \frac{l}{2}}e^{- \frac{(x - u_{\text{x}} \cdot T_k)^2}{4D_{x}T_k}}\,dx,
\label{eq:integral}
\end{equation}
and $x_r$ is the receiver's location on $x$, $l$ is the edge length of cube-shaped volume to be absorbed, $u_x$ is the average flow rate on the $x$ direction, $T_k$ is given in (\ref{eq:absorptionTimeK}), $D_x$ is the diffusion coefficient on the $x$ direction.

After calculating the integral in (\ref{eq:integral}) using substitution method, we get
\begin{equation}
\begin{aligned}
I_x &= \sqrt{\pi D_x T_k}\\&\Bigg[ \mathsmaller{   \text{erf}\left(\frac{x_r + \frac{l}{2} - u_{\text{x}}(n+1-k)T_k}{\sqrt{4D_xT_k}}\right) - \text{erf}\left(\frac{x_r - \frac{l}{2} - u_{\text{x}}(n+1-k)T_k}{\sqrt{4D_xT_k}}\right)      }   \Bigg],
\end{aligned}
\label{eq:ix}
\end{equation}
where
\begin{equation}
\text{erf}(z) = \frac{2}{\sqrt{\pi}} \int_{0}^{z} e^{-t^2} \, dt.
\end{equation}
$I_y$ and $I_z$  can also be calculated similarly due to symmetry. As a result, $M_{a_n}$ can be expressed as
\begin{equation}
\begin{aligned}
  M_{a_n} = \sum_{k=1}^{n} \frac{M_{n+1-k}}{8} \cdot I_x \cdot I_y \cdot I_z,
\end{aligned}
\label{eq:mn}
\end{equation}
where $I_x$ is given in (\ref{eq:ix}). 

Total absorbed mass in the $n^{th}$ symbol given by (\ref{eq:mn}), consist of mass $M_n$ that is released for $n^{th}$ symbol and masses $M_{n-1},..., M_1$ that are released for previous symbol intervals. The coefficients in front of the masses indicate the proportion of the released mass that is being absorbed. For receiver to perform next stages with less error, we would like to absorb at the time when $M_n$ reaches its peak value. To maximize $M_n$, we need the maximize the coefficient of $M_{n+1-k}$ in (\ref{eq:mn}) when $k = 1$ and to maximize the coefficient, we need to maximize $I_x \cdot I_y \cdot I_z$, with respect to time.  Time variable of these integrals is $T_k$. When we substitute $k = 1$ in (\ref{eq:absorptionTimeK}), we get $T_1 = T_a$. Hence, we need to maximize the multiples of integral with respect to absorption time in a symbol interval: $T_a$. For this, we can utilize numerical methods within a software environment such as MATLAB.

After finding $T_a$, it is also required to decide on $T_s$. We can set a relation between $T_s$ and  $T_a$ as $T_s = m \cdot T_a$. In this relation, choice of $m$ affects the reliability and data rate of the system. Whilst higher choices for $m$ reduce the ISI, they also reduce the data rate. So, $m$ becomes an important system parameter. We name this parameter as the symbol to sampling ratio.

\begin{figure*}[t]
    \centering
    \begin{subfigure}{0.245\linewidth}
        \includegraphics[width=\linewidth]{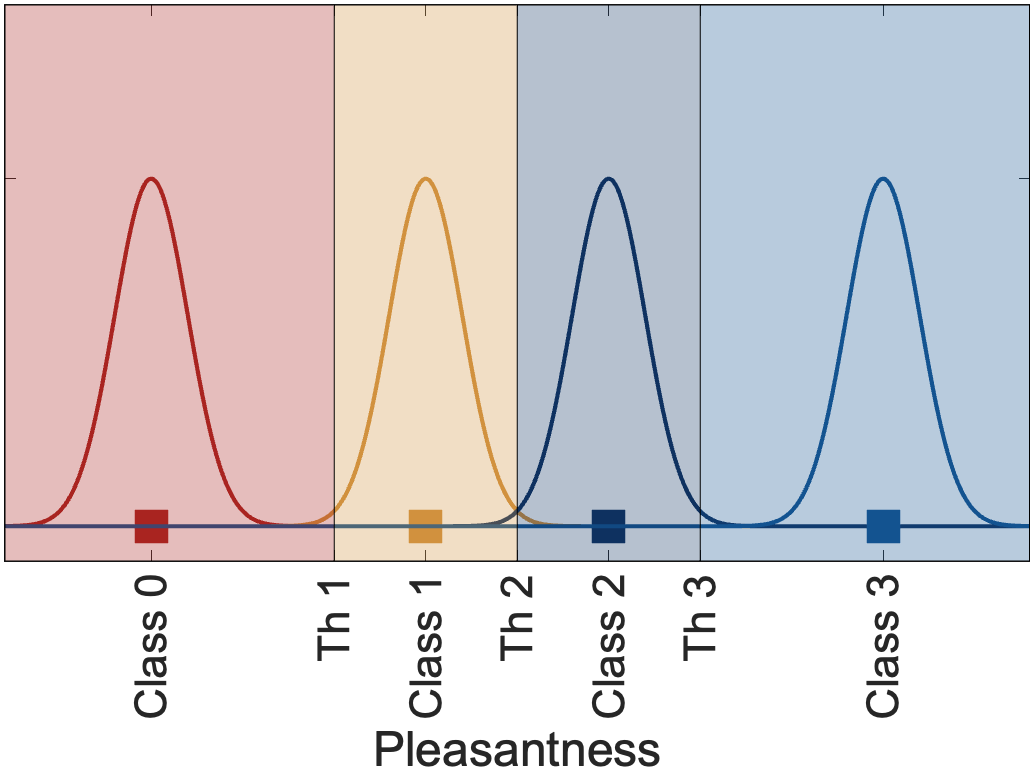}
        \caption{ $Q_p:1, PN:5$}
        \label{fig:sub1}
    \end{subfigure}
    \begin{subfigure}{0.245\linewidth}
        \includegraphics[width=\linewidth]{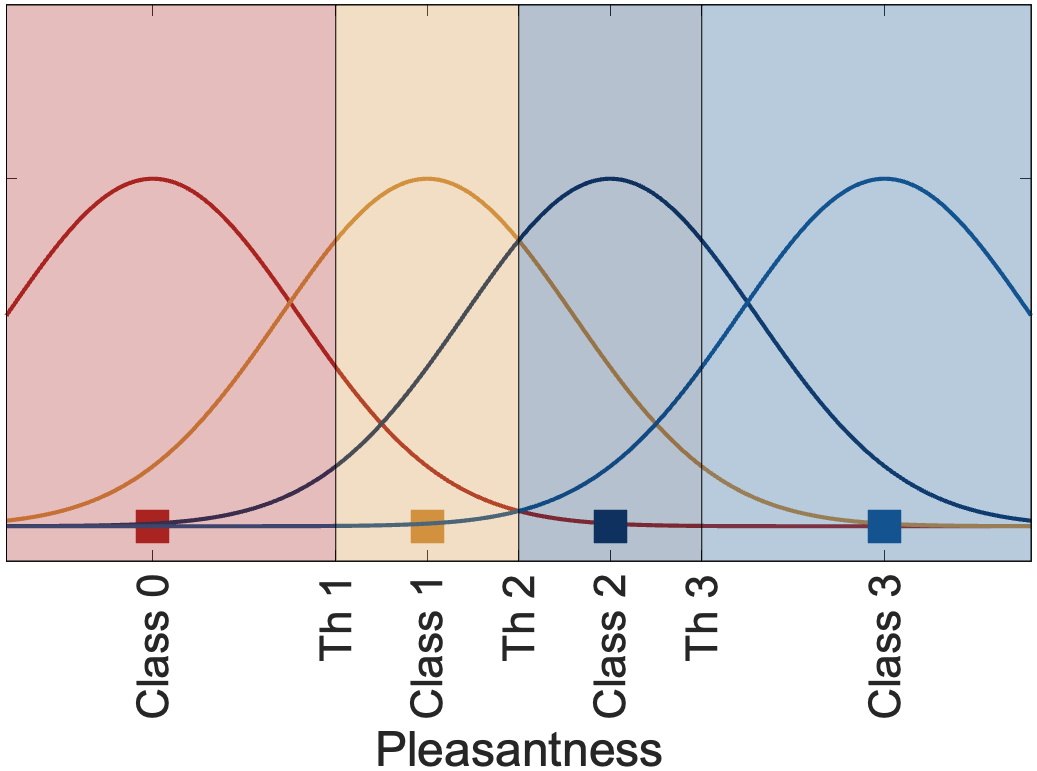}
        \caption{ $Q_p:1, PN:20$}
        \label{fig:sub2}
    \end{subfigure}
    \begin{subfigure}{0.245\linewidth}
        \includegraphics[width=\linewidth]{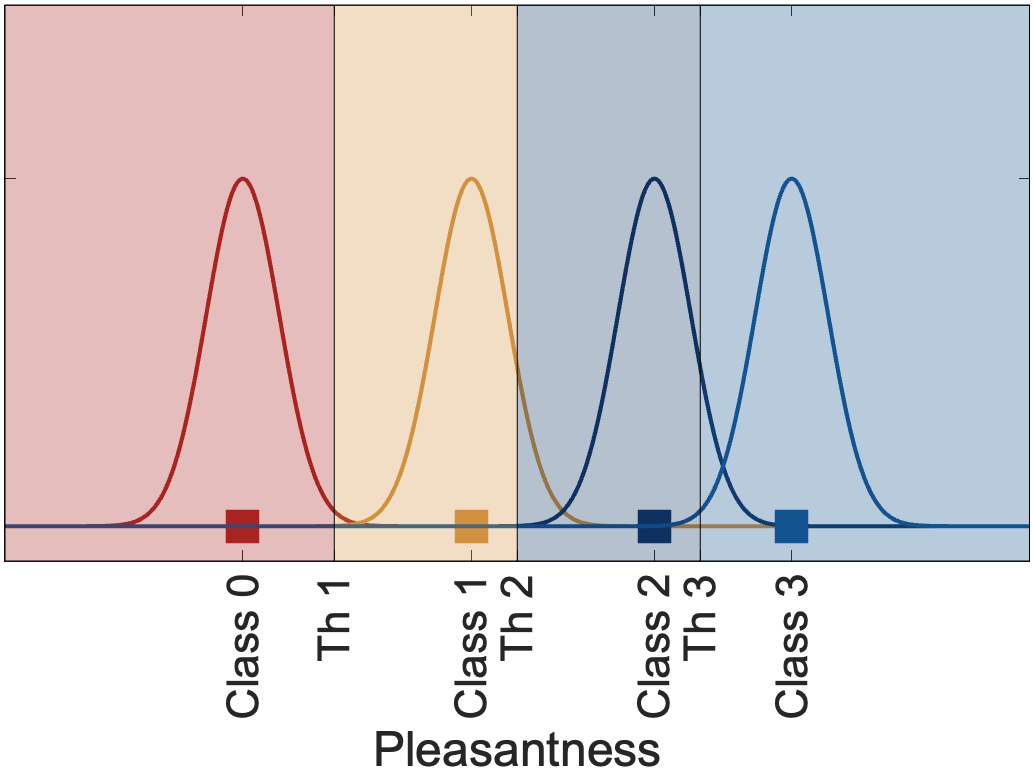}
        \caption{ $Q_p:0.5, PN:5$}
        \label{fig:sub3}
    \end{subfigure}
    \begin{subfigure}{0.245\linewidth}
        \includegraphics[width=\linewidth]{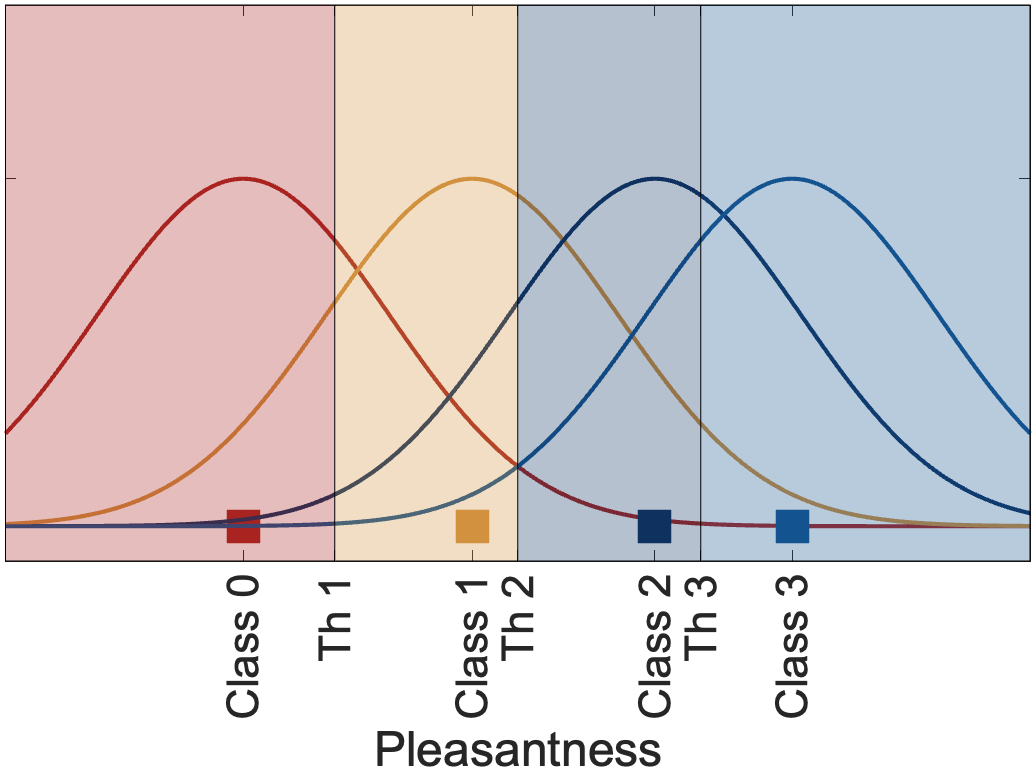}
        \caption{ $Q_p:0.5, PN:20$}
        \label{fig:sub4}
    \end{subfigure}
    \caption{The pleasantness values of odors at odor bank marked as squares. Gaussian curves represent the demodulation output for pleasantness, denoted as $p_{received}$. The areas between threshold values are shaded to indicate the decision regions.}
    \label{fig:odorQualityGraphs}
\end{figure*}

\subsubsection{Demodulation}\label{Receiver:Demodulation}
Following the absorption process, receiver selects the odor with the greatest mass within the absorbed sample and calculates the selected odor's perceptual value vector $(p_{received}, i_{received}, e_{received})$ with its processor unit. Examples of this processor can be find in \cite{enosepleasantness}.

Under ideal conditions, we would expect selected odor's perceptual value vector, $(p_{received}, i_{received}, e_{received})$, to be exactly the same with the transmitted odor's perceptual value vector : $(p_o, i_o, e_o)$, see section \ref{section:Transmitter}. However, under practical conditions, calculation of perceptual value vector can not perfectly achieved at the processor unit. This introduces a noise to system. Let's name this specific type of noise within the receiver as processor noise and denote it by $PN$. 

In \cite{gaussian}, processor calculations are modelled as Gaussian random variables. Therefore, we also assume $(p_{received}, i_{received}, e_{received})$ values are Gaussian random variables with means $(p_o, i_o, e_o)$ and a standard deviation ${\sigma_{processor}}$ for each dimension. The standard deviation of these random variables serves as a measure of $PN$, as expressed by
\begin{equation}
    PN = \sigma_{processor}. 
    \label{eq:qualityRelationBetweenDimensions}
\end{equation}

In the decoding stage, which follows the demodulation stage, the receiver identifies the odor class of the selected odor from the absorbed sample by utilizing the perceptual value vector calculated in the processor unit. A decoding error occurs when the calculated odor's perceptual value vector does not fall within the expected region of the perceptual vector space. Consequently, deviations in calculating the selected odor's perceptual value vector may lead to decoding errors in the subsequent stage. Higher $PN$ values results in spreader Gaussian curves along the perceptual axes, increasing the likelihood of decoding errors. Assuming the areas between threshold values as different decision regions, Fig. \ref{fig:odorQualityGraphs} illustrates the effect of processor noise by comparing two distinct processor noise levels: one at $PN = 5$ and the other at $PN = 20$. As expected, lower $PN$ value results in a more centralized Gaussian curve around the mean, reducing the likelihood of crossing the borders of decision regions. Conversely, higher $PN$ value leads to a spreader Gaussian curve, increasing the risk of crossing the borders of decision regions.

As it may also be realized, another factor influencing the likelihood of crossing the decision region boundaries within the perceptual vector space is the location of the mean within the decision region. The mean locations are related to the perceptual value vectors of the odors selected to initialize the odor bank. During the initialization of the odor bank, an odor is selected for each class, as discussed in section \ref{section:Transmitter}. Given the vast array of odors in nature, a single odor class can be represented by many different odors, each with perceptual value vectors that fall within the class's region in the odor perceptual space. Nonetheless, certain odor selections perform better in minimizing decoding errors.

To elucidate the impact of selecting an odor set for the odor bank, we introduce a system parameter that is called quality of the odor set, denoted by $Q$. Before mathematically defining the quality of an odor set, let's first understand the rationale behind this concept. To minimize decoding errors, it is desirable for the Gaussian curves to have shorter tails that do extend into the neighbor regions. This can be achieved either by reducing processor noise or by positioning the means closer to the center of the decision regions. This implies that as the means move closer to the decision region boundaries, the tails of the Gaussian curves extend further into the neighbor decision regions. Consequently, odor sets comprising odors positioned near the central locations of decision regions can be considered to have higher odor set quality. Therefore, in the mathematical formulation of quality, the distances from the optimal odor locations are penalized. The optimal locations are identified as the central points for inner regions and the boundary points for outer regions. With this reasoning, we can now mathematically define the quality of an odor set.

The smallest unit of our system is the odor that represent group of $K(n_p-n_i-n_e)$ bits. Thus, any error occurs in the reproduction of $n_p$, $n_i$, or $n_e$ bits, will result in a symbol error. Regardless of the qualities of other two dimensions, if the quality of one dimension is not good then symbol error will be dominated by errors on the reproduction of bits belong to this dimension. Therefore, we derive overall quality from individual quality of perceptual dimensions as
\begin{equation}
    Q = min(Q_p, Q_i, Q_e), 
    \label{eq:qualityRelationBetweenDimensions}
\end{equation}
where $Q_p$, $Q_i$, and $Q_e$ be the quality of pleasantness, intensity, and edibility, respectively. Then we define the quality of pleasantness as
\begin{equation}
    \begin{aligned}
    Q_p &= 1 - \frac{|p_{O_o} - 0|}{\frac{100}{2^{n_p}}} + \frac{|100 - p_{O_{2^{n_p}-1}}|}{\frac{100}{2^{n_p}}} \\&+ \sum_{k = 1}^{2^{n_p} - 2} \mathsmaller{  \frac{||p_{O_k} - max(p_T : p_T < p_{O_k})| - |min(p_T : p_T > p_{O_k}) - p_{O_k}||}{\frac{100}{2^{n_p}}}   },
    \end{aligned}
    \label{eq:qualityDefinitionForPerceptualAxes}
\end{equation}
where $p_{O_k}$ represent the pleasantness value of odor molecule that represents the pleasantness class $k$ such that $k \in [0, 2^{n_p} - 1]$ and $p_T$ is the threshold value set that is defined in (\ref{eq:pt}). $Q_i$, and $Q_e$ can also be expressed similarly. 

Fig. \ref{fig:odorQualityGraphs} also demonstrates the influence of the quality parameter by maintaining constant processor noise levels at $PN = 5$ and $PN = 20$. The graphs reveal that, for the same processor noise, higher quality sets exhibit a lower probability of crossing the decision region boundaries.

\subsubsection{Decoding}\label{Receiver:Decoding}
In the final decoding phase, we need to determine the pleasantness, intensity and edibility classes $(p, i, e)$ of the received symbol by utilizing the processor unit's output $(p_{received}, i_{received}, e_{received})$. For this, we can use (\ref{eq:perceptual_code}). Then, we can convert decimal values to their binary values which gives $n_p$, $n_i$, and $n_e$ bits for corresponding dimensions. As a result, $n_p$, $n_i$, and $n_e$ bits construct the block of $K$ bits that was targeted to be transmitted for this symbol interval.

\section{Adaptive Symbol Transmission}

In \ref{section:Transmitter}, it is assumed that the odor bank on the transmitter side consists of $2^{n_p} \cdot 2^{n_i} \cdot 2^{n_e}$ different odor classes. Furthermore, each odor class is matched with an odor whose perceptual value vector falls within the region of the corresponding class. However, it can be argued that over time, a specific type of odor may deplete due to the high frequency of the bit sequence represented by this odor. This might cause a shorter operation time for systems which are hard to refill with depleted odor capsules.

In this section, to overcome this problem, we propose an adaptive transmitter and receiver. Depending on the depleting odor type, number of bits represented by each perceptual dimension can be updated from $K(n_p,n_i,n_e)$  to $K^{'}({n_p}^{'},{n_i}^{'},{n_e}^{'})$.  Hence, operation time of system can be extended.

This approach adds complexity to the system. Thus, it can also be argued whether it is worth adding this complexity to extend the system's operation time. To address this question, transmitter estimates the anticipated extension in the system's operation time. Then, if updating the system extends the operation time by $E$, where $E$ is the system parameter determined according to the cost of complexity, then it proceeds with the system update.

On the other hand, the receiver is responsible for detecting the system update and adjusting its decoding stage based on the received update information.

\begin{algorithm}[!t]
\begin{algorithmic}[1]

{
\ttfamily
\State \textbf{Start}
\While{true}
    \If{transmitted symbols mod $N$ == 0}
        \If{can send $N$ more symbols}
            \State Continue to transmit symbols
        \Else
            \State Enter to update state
            \State $C=[(n_p^{'}, n_i, n_e), (n_p, n_i^{'}, n_e), (n_p, n_i, n_e^{'})]$
            \State Initialize $T \gets \{\}$ and $i \gets 1$
            \While{$i \leq \text{length}(C)$}
                \State Estimate $T_i$
                \State Append $T_i$ to $T$
                \State $i \gets i + 1$
            \EndWhile
            \State Set $T_{max} = max(T)$
            \State Set $i_{max} = argmax(T)$
            \If{ $T_{max} \geq E$}
                \State Re-init tx with $C[i_{max}]$
                \State Inform rx about the update
                \State Continue to transmit symbols
            \Else
                \State \textbf{End}
            \EndIf
        \EndIf
    \Else
        \State Continue to transmit symbols
    \EndIf
\EndWhile
}

\end{algorithmic}
\caption{Adaptive symbol transmission algorithm.}
\label{algorithm}
\end{algorithm}

\subsection{Transmitter}

For the system update, the transmitter undertakes a few tasks. Firstly, it determines the time when to implement system update. For this, at every $N$ symbol transmission, the transmitter checks the statistics of the last $N$ symbols. Based on this data, if there is insufficient remaining odor molecules to transmit the next $N$ symbol, the transmitter goes into update state. 

In the update state, transmitter needs to decide on how to update number of bits represented by each perceptual dimension: $K(n_p,n_i,n_e)$. There exists three possible combinations: $(K-1)(n_p-1,n_i,n_e)$, $(K-1)(n_p,n_i-1,n_e)$, $(K-1)(n_p,n_i,n_e-1)$. Transmitter simulates each combination.

Simulations start by reassigning odor class codes to the remaining odor capsules in the odor bank, similar to what is described in \ref{Transmitter:Initialization}. However, this time, there is a slight distinction: some odor classes can be represented by more than one odor since number of regions in the perceptual space change from $2^{K}$ to $2^{K-1}$. This slight distinction positively impacts the system by increasing the odor options for certain odor classes in the new system. Subsequently, transmitter anticipates the possible extension in the system's operation time based on the data of last $N$ symbol transmission.

After simulations, transmitter picks the combination with the highest extension time. Then it compares this value with the system parameter $E$. If it is higher than the system parameter, then transmitter starts the system update for the picked combination. System update starts by the re-initialization of transmitter. After re-initialization, transmitter informs the receiver about the system update. This involves three things: first, letting receiver know there is a change in the system, second, explaining how receiver should decode odors from now on, and third letting receiver know the system update ended.

Transmitter can inform the receiver about the change by refraining from transmitting anything for a certain number of symbol intervals. During the silence in the channel, receiver detects that there is going to be a system update. However, since receiver still does not know the new combination, it will decode the incoming odor after silence according to combination before update: $K(n_p,n_i,n_e)$. Therefore, transmitter needs to send the information of new combination by using the old combination and old odor class codes. Hence, transmitter sends the odor which was representative of the odor class $O_{{n_p}^{'}{n_i}^{'}{n_e}^{'}}$. By this means, pleasantness class of the decoded odor will be interpreted by receiver as the number of bits that is going to be represented by pleasantness dimension in the new combination. This interpretation is the same for other dimensions as well. At the end of this transmission, receiver becomes informed about the new combination: $K^{'}({n_p}^{'},{n_i}^{'},{n_e}^{'})$.  

Since the system update is critical, we might prefer to send the odor carrying information about the new combination for a few symbol intervals. This can be seen as a precaution against symbol errors that may occur in the initial transmission. Thanks to the repeated transmission of the odor, the receiver selects the one that occurs most frequently in the received sequence after the period of silence. 

To inform the receiver about the end of system update, transmitter again goes into a silence mode for a certain number of symbol intervals. After this last silence, system continues to transmit information with the new combination. Algorithm \ref{algorithm} displays the pseudo code of the algorithm that is being explained.

\subsection{Receiver}

As we explained in \ref{Receiver:Demodulation}, our receiver selects the odor with the greatest mass within the absorbed sample for each symbol interval. In this case, even if transmitter does not send anything for a symbol interval, the leftover molecules from the transmission of previous symbols could be detected as a new symbol that is being transmitted. That is why receiver needs to be equipped with the silence detection feature. 

If we set $M_T$ as the minimum mass required for the greatest mass within the absorbed sample, we can identify periods of silence in the channel. In this scenario, $M_T$ should be determined based on the released mass for each symbol, M, and the expected mass ratio of the system discussed in \ref{Section:Mass Ratio Analysis}. 

\begin{figure*}[t]
    \centering
    \begin{subfigure}{0.245\linewidth}
        \includegraphics[width=\linewidth]{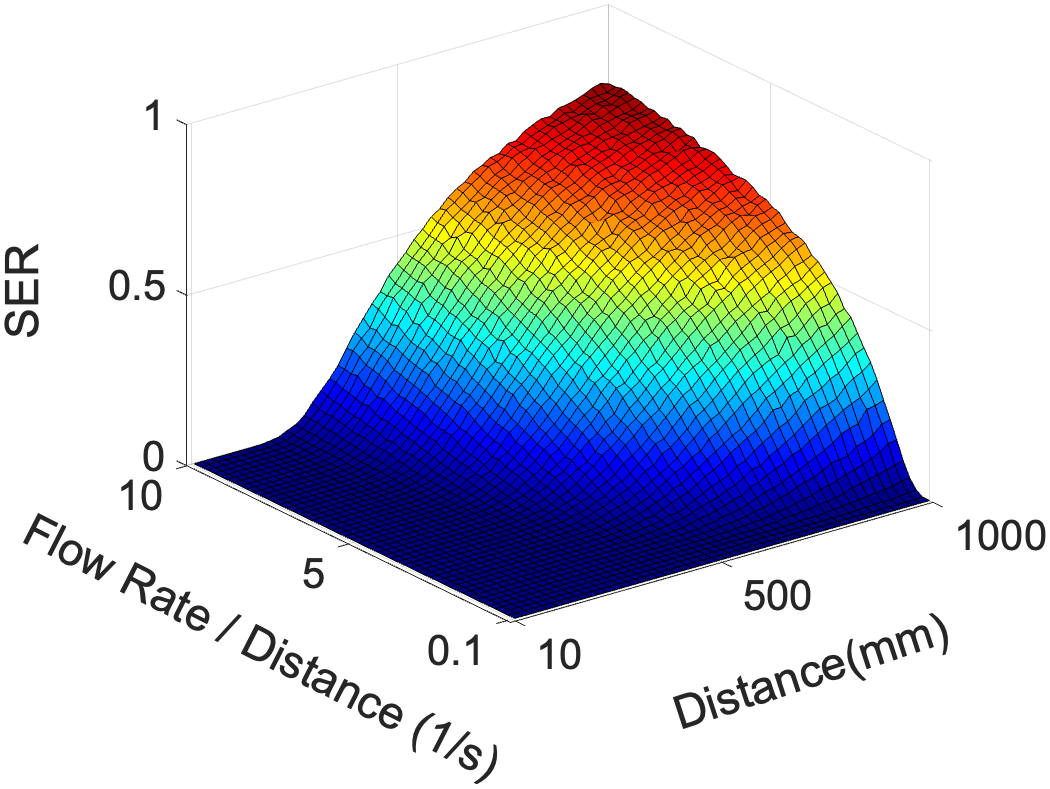}
        \caption{edge / distance = 0.001}
        \label{fig:ser_sub1}
    \end{subfigure}
    \begin{subfigure}{0.245\linewidth}
        \includegraphics[width=\linewidth]{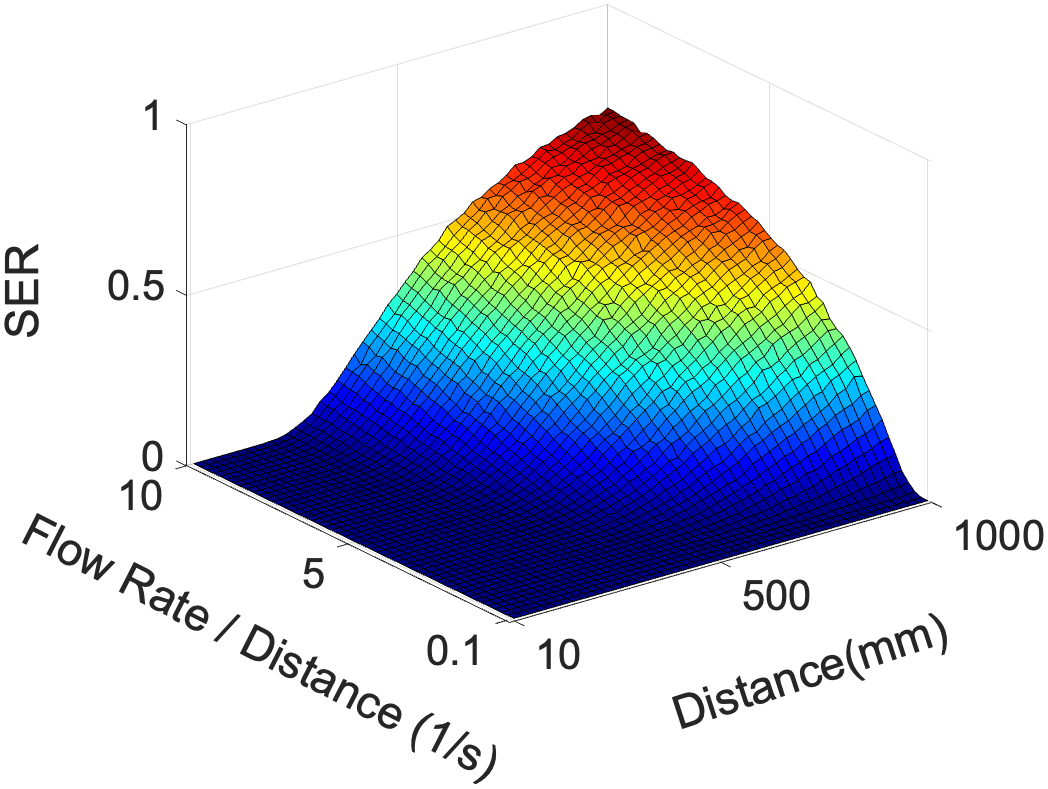}
        \caption{edge / distance = 0.01}
        \label{fig:ser_sub2}
    \end{subfigure}  
    \begin{subfigure}{0.245\linewidth}
        \includegraphics[width=\linewidth]{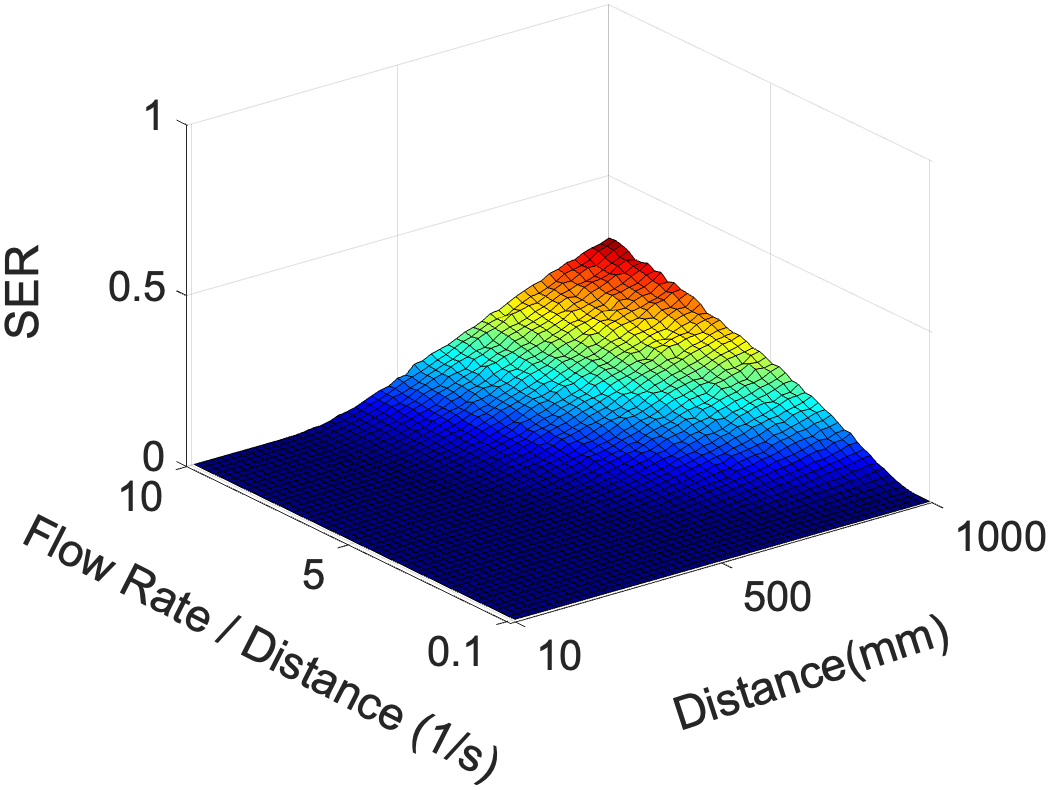}
        \caption{edge / distance = 0.05}
        \label{fig:ser_sub3}
    \end{subfigure}
    \begin{subfigure}{0.245\linewidth}
        \includegraphics[width=\linewidth]{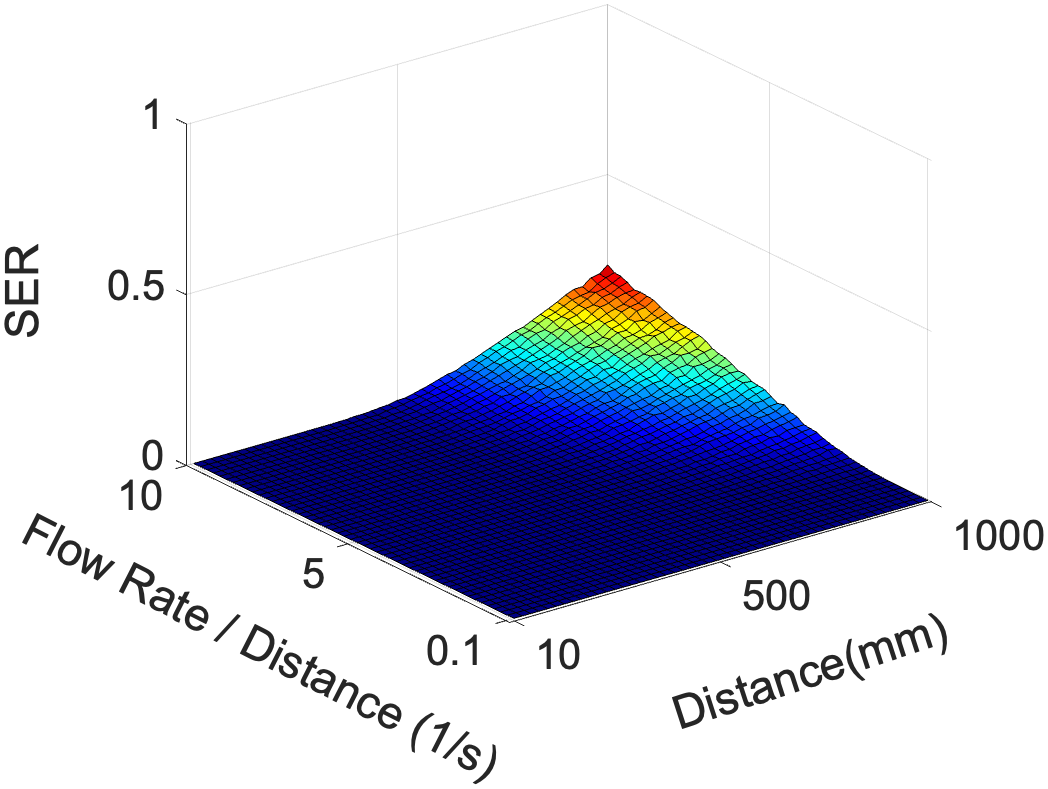}
        \caption{edge / distance = 0.1}
        \label{fig:ser_sub4}
    \end{subfigure}
    \caption{$SER$ Analysis for different edge length to transmitter-receiver distance ratios. ${(FNR_x, FNR_y, FNR_z) = (20, 20, 20)}$}
    \label{fig:type1SERAnalysis_flowrate_distance_edgeLength}
\end{figure*}
\section{Performance Evaluation}

In this section, we present the analysis results of the odor-based molecular communication system implemented using the OPSK modulation method. To achieve this, we employed a mathematical model constructed in MATLAB, based on the theory outlined in the previous chapters. We investigated the effect of system parameters like edge length of cube-shaped absorption volume, flow rate of the channel, distance from transmitter to receiver, noise in the channel, processor noise $(PN)$, and quality of the odor set $(Q)$ on performance parameters like symbol error rate $(SER)$, symbol rate $(SR)$ and mass ratio. For analyses, we created a random symbol sequence of length 10000, we assumed that symbol to sampling ratio, $m$, is equal to 2, each odor molecule has the constant diffusion coefficient $D = 0.14\cdot10^{-4} \frac{m^2}{s}$ at every direction, and the released odor mass at each symbol is $M = 2.4\cdot10^{-9}$ kg.

For some analyses, we investigated the effects in 3D plots to observe the effect of multiple parameters at the same time. To improve the outcome of analyses, we scaled edge length of cube and channel flow rate with the corresponding distances. Therefore, we redefined the edge length and flow rate as the ratio of edge length to distance and ratio of flow rate to distance.

In our concluding analysis, we explored the impact of adaptive symbol transmission on extending operation time. To do this, we examined configurations with different number of bits per symbol. For each configuration, we generated different frequency distributions of symbols in the data to be transmitted.

\begin{figure*}[t]
    \centering
    \begin{subfigure}{0.245\linewidth}
        \includegraphics[width=\linewidth]{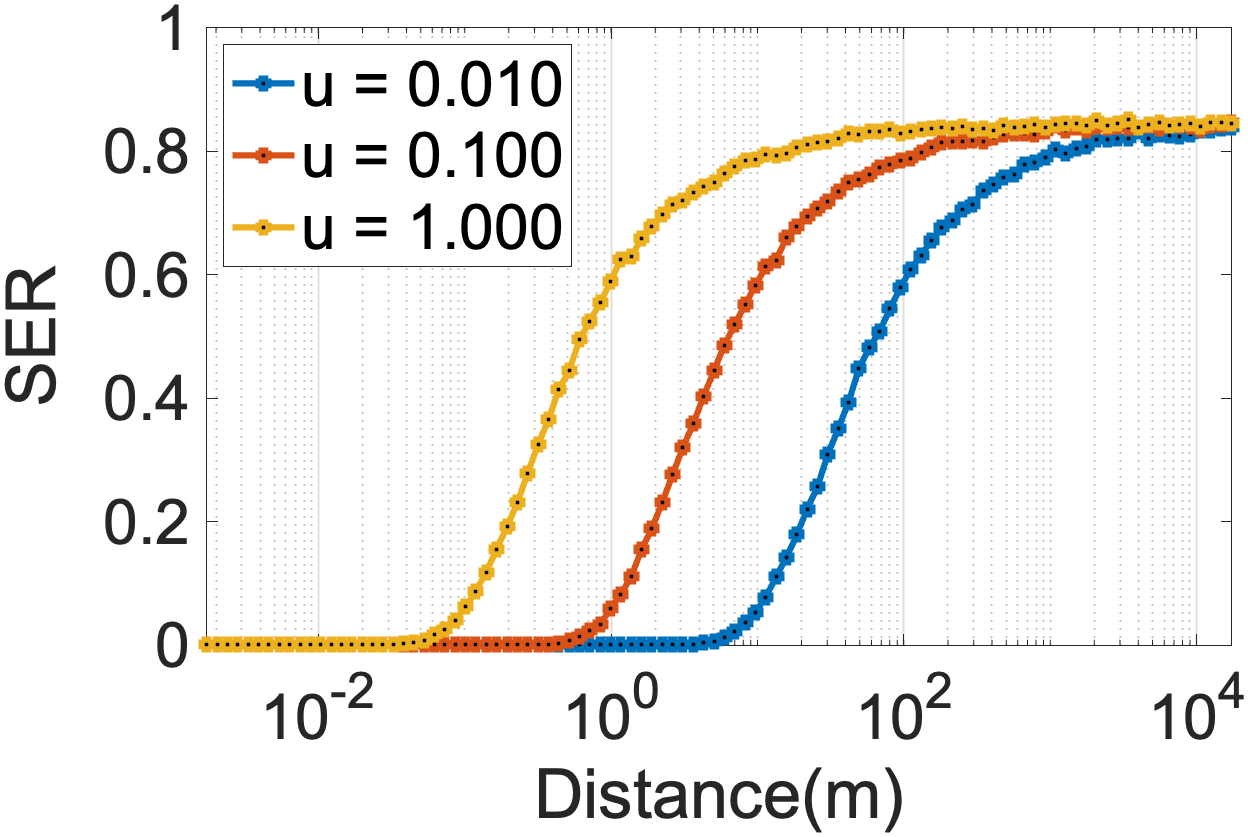}
        \caption{$FNR = 10$}
        \label{fig:dist_sub1}
    \end{subfigure}
    \begin{subfigure}{0.245\linewidth}
        \includegraphics[width=\linewidth]{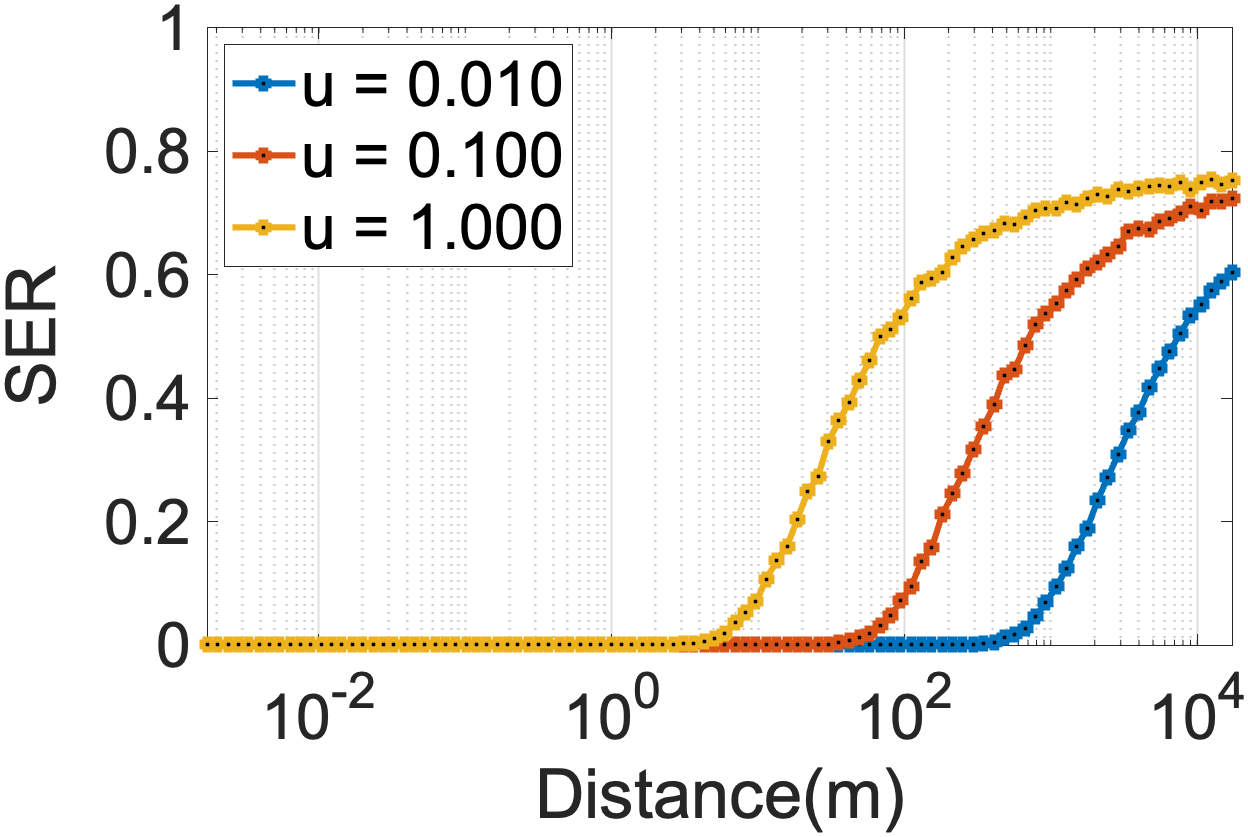}
        \caption{$FNR = 50$}
        \label{fig:dist_sub2}
    \end{subfigure}  
    \begin{subfigure}{0.245\linewidth}
        \includegraphics[width=\linewidth]{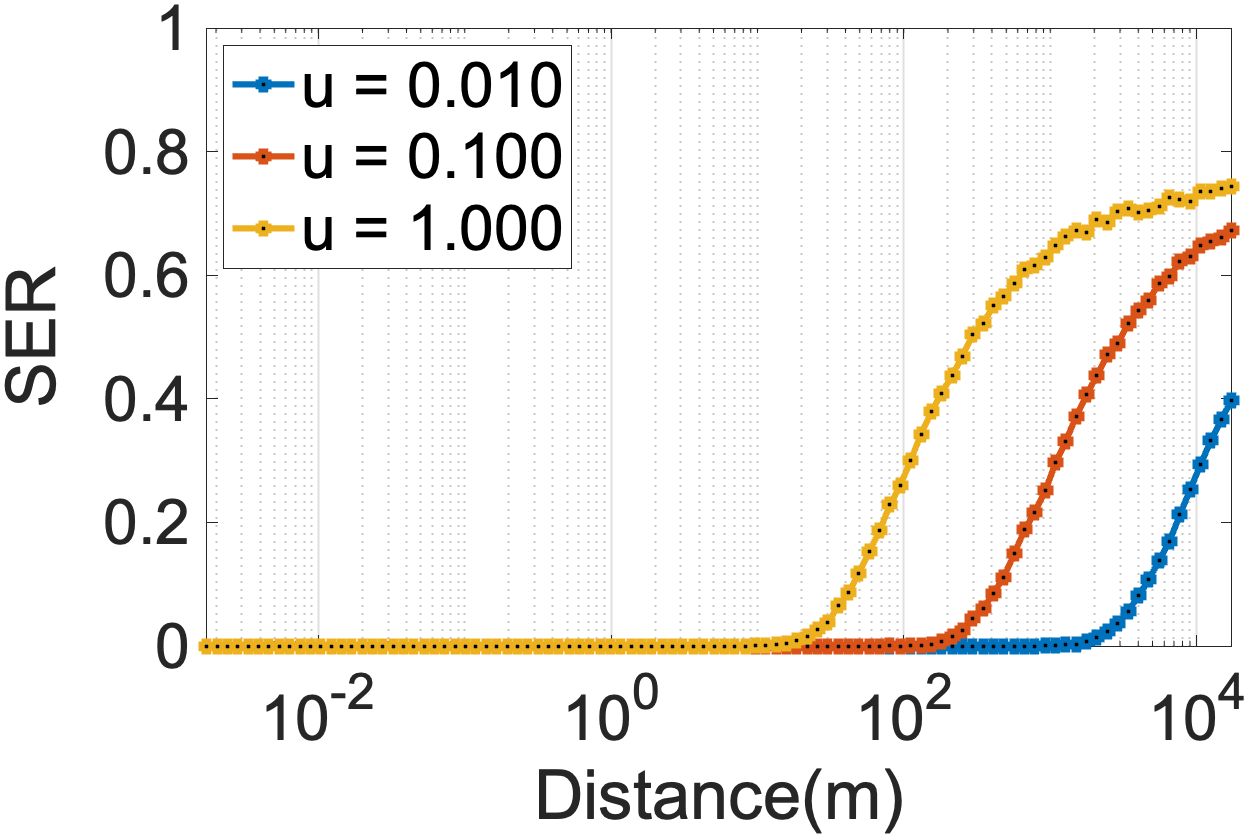}
        \caption{$FNR = 100$}
        \label{fig:dist_sub3}
    \end{subfigure}
    \begin{subfigure}{0.245\linewidth}
        \includegraphics[width=\linewidth]{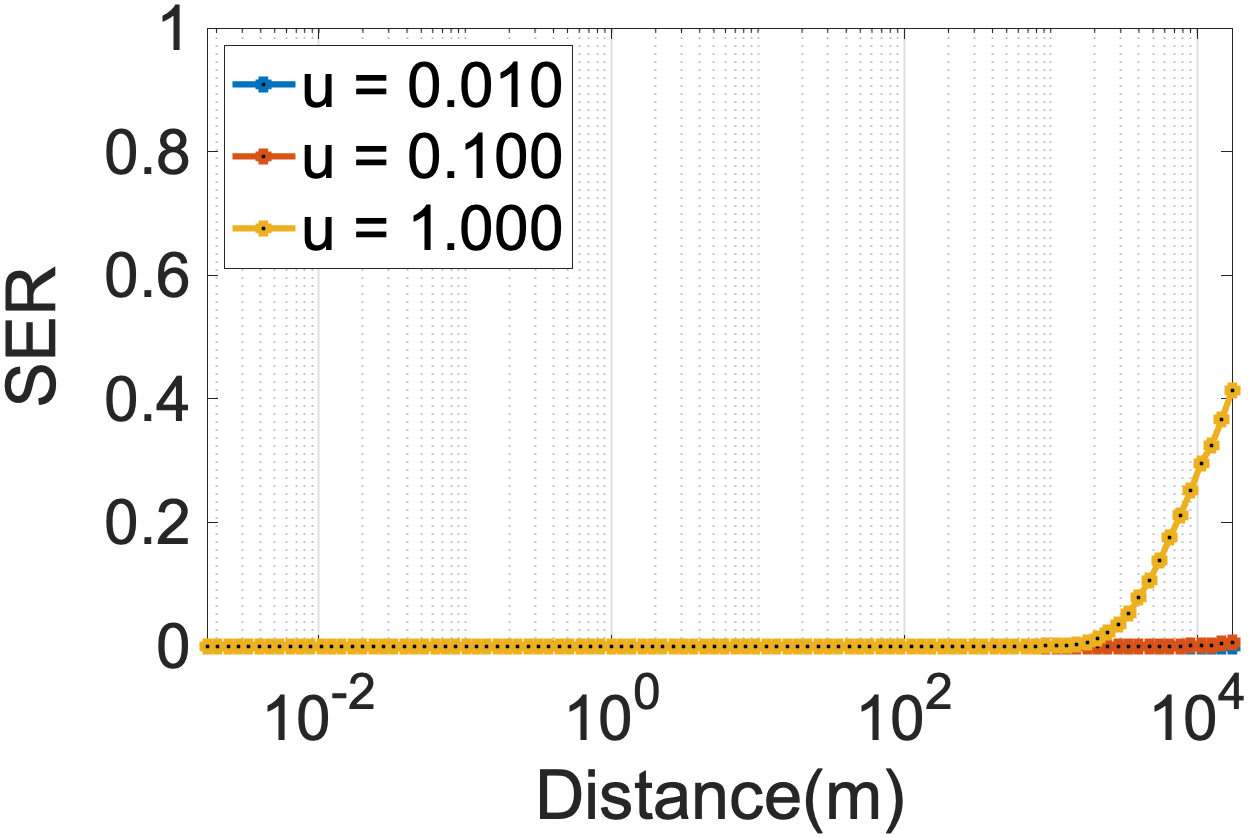}
        \caption{$FNR = 1000$}
        \label{fig:dist_sub4}
    \end{subfigure}
    \caption{Effect of distance on type 1 $SER$ for different flow rates and $FNR$ values where ${(FNR_x, FNR_y, FNR_z) = (FNR, FNR, FNR)}$.}
    \label{fig:distance_analysis}
\end{figure*}
\begin{figure*}[t]
    \centering
    \begin{subfigure}{0.245\linewidth}
        \includegraphics[width=\linewidth]{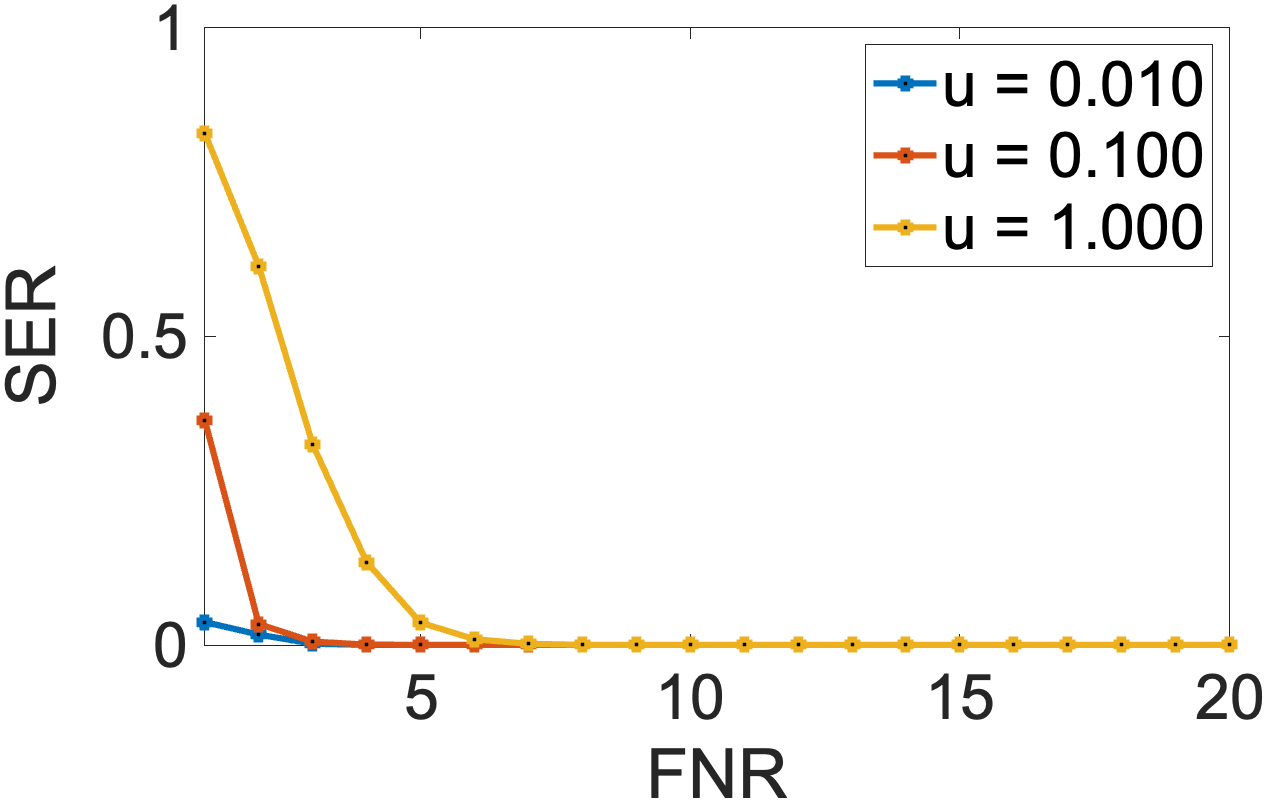}
        \caption{$distance = 1 cm$}
    \end{subfigure}
    \begin{subfigure}{0.245\linewidth}
        \includegraphics[width=\linewidth]{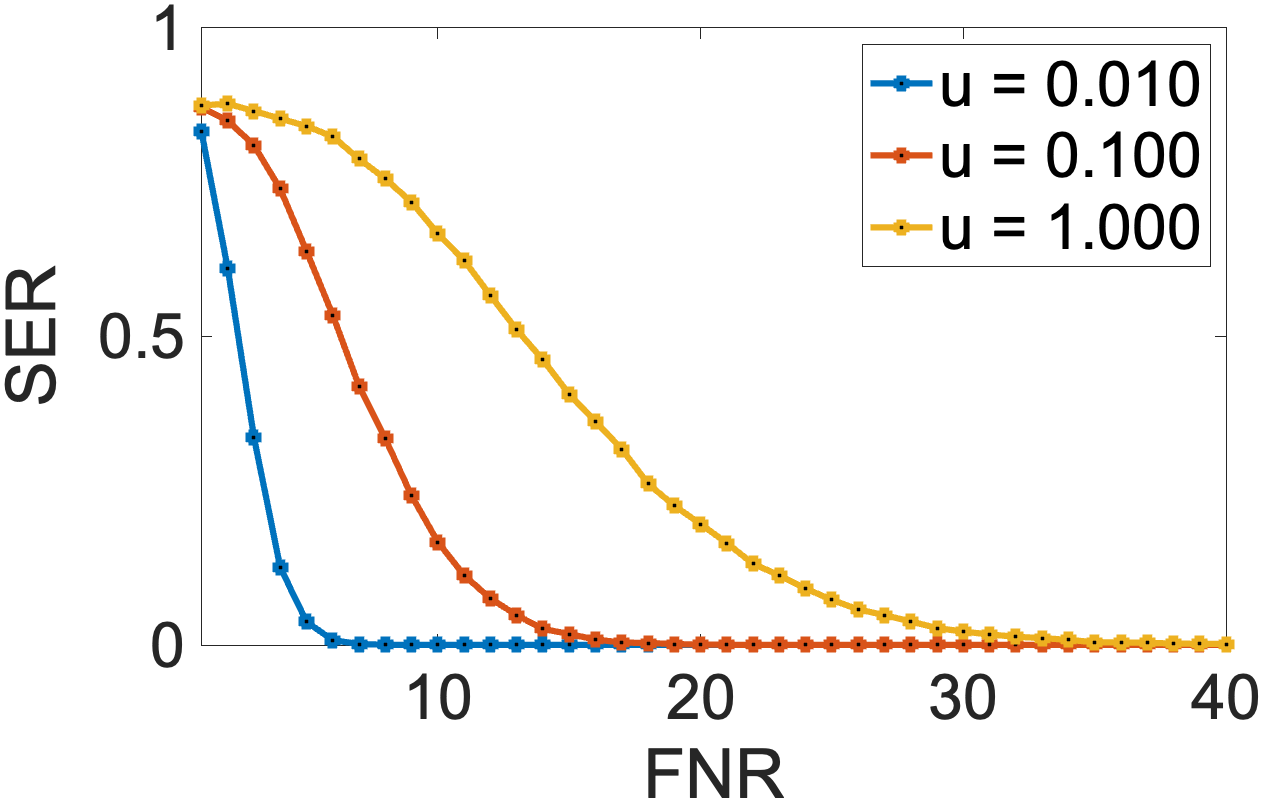}
        \caption{$distance = 1 m$}
    \end{subfigure}  
    \begin{subfigure}{0.245\linewidth}
        \includegraphics[width=\linewidth]{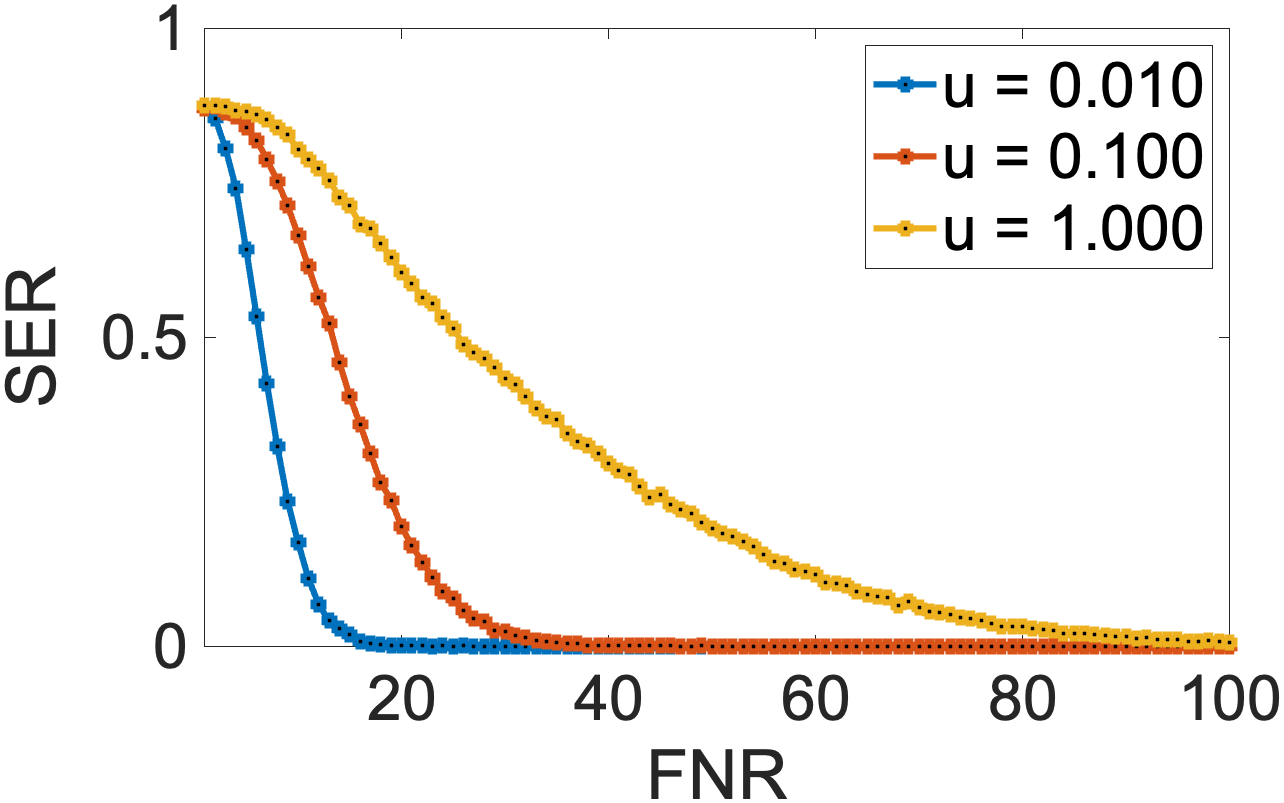}
        \caption{$distance = 500 m$}
    \end{subfigure}
    \begin{subfigure}{0.245\linewidth}
        \includegraphics[width=\linewidth]{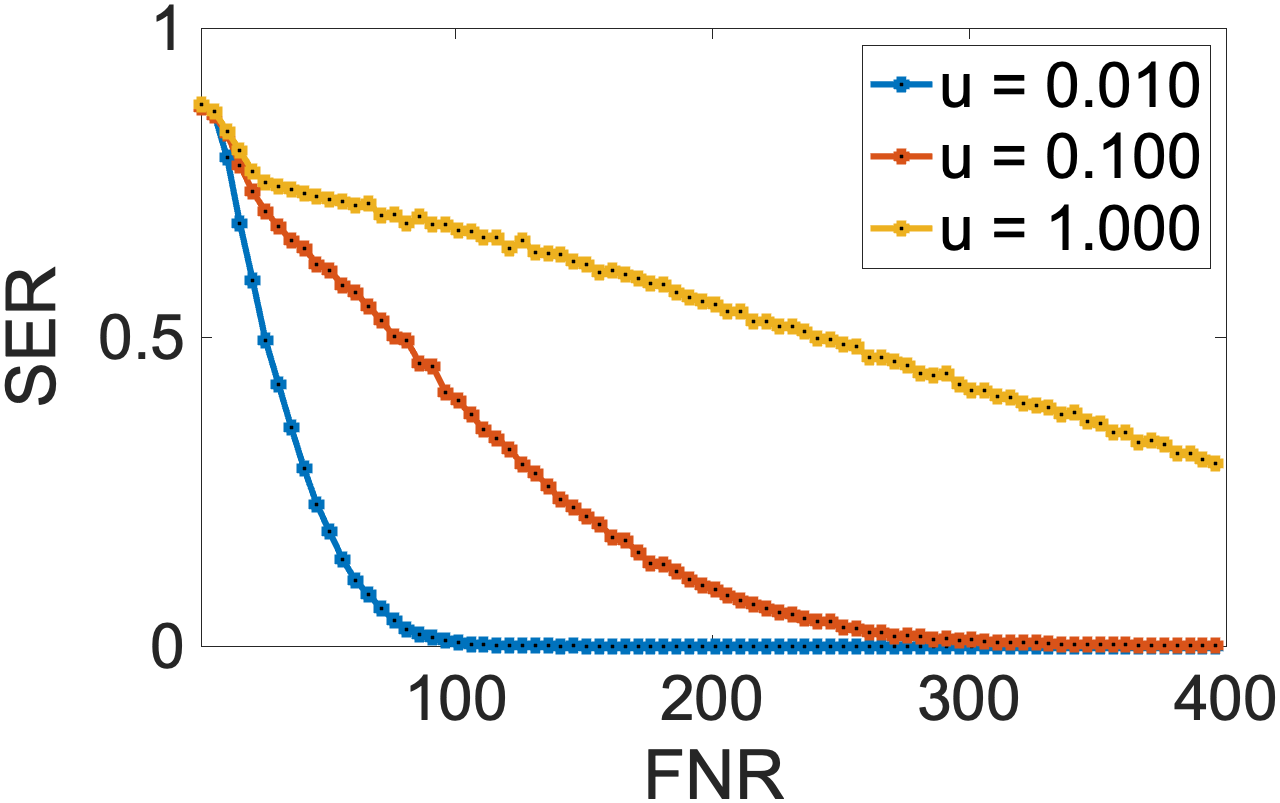}
        \caption{$distance = 1 km$}
    \end{subfigure}
    \caption{Effect of $FNR$ where ${(FNR_x, FNR_y, FNR_z) = (FNR, FNR, FNR)}$ on type 1 $SER$ for different distances and flow rates.}
    \label{fig:fnr_analysis}
\end{figure*}
\subsection{$SER$ Analysis}
As outlined in section \ref{section:Receiver}, the receiver processes the absorbed sample through two operations, each impacting the $SER$ differently. To distinguish the effects of two operation on the overall system performance we separate their analyses.  

The initial operation at the receiver side involves selecting the odor with the greatest mass within the absorbed sample. If the sample contains other odors with greater mass than the one transmitted for the most recent symbol, this leads to incorrect odor selection, thereby contributing to $SER$. The reason of incorrect odor selection may be attributed to various system parameters, including flow rate, the ratio of edge length to distance, distance itself, and $FNR$. Such incorrect selections result in symbol errors, which we will refer to as type 1 errors.

The second operation at the receiver side involves calculating the selected odor’s perceptual value vector and decoding accordingly. Following the demodulation stage, the receiver attempts to identify the region of the selected odor within the perceptual vector space. Decoding errors arise when the perceptual value vector for the selected odor is positioned in an incorrect decision region. Higher $PN$ values of processor and lower $Q$ values of odor set are factors contributing to symbol errors, which we will refer to as type 2 errors. 

For the sake of investigating the effects of these two operations to overall system separately, we assumed noise-free processor while simulating the effect of first operation and noise-free channel while simulating the effect of second operation. 

\subsubsection{Type 1 $SER$ Analysis}
Fig. \ref{fig:type1SERAnalysis_flowrate_distance_edgeLength}, displays the analysis results at each subplot for different edge length to distance ratios. Each subplot has distance ranges from $10\hspace{1mm}mm$ to $1\hspace{1mm}m$ on the x-axis, flow rate to distance ratio ranges from $0.1$ to $10$ on the y-axis and $SER$ on the z-axis. Displayed results are for $(FNR_x, FNR_y, FNR_z) = (20, 20, 20)$. 

When we evaluate plots all together, we observe that increasing edge length to distance ratio gradually from $0.001$ to $0.1$ increases the amount of lastly released odor mass at the absorbed sample. Hereby, correct odor’s selection chance increases which result in decrease on $SER$. 

Wrong odor selection probability increases with the increasing flow rate because the propagation of molecules in the channel more dominated by the flow rate rather than diffusion at higher flow rates. Therefore, we observe higher $SER$ values at higher flow rate values and lower $SER$ values at lower flow rate values. 

In Fig. \ref{fig:type1SERAnalysis_flowrate_distance_edgeLength}, when we keep the flow rate to distance ratio constant and increase the distance, we implicitly increase the flow rate. That’s why, the behavior in the x-axis is similar to behavior in the y-axis for the same reasons. Therefore, to see the effect of distance separately we did another analysis. 

Fig. \ref{fig:distance_analysis} displays the results of distance analysis. Previous analysis already show that low $SER$ values can be obtained in several cases. Therefore, in this analysis we tested the long-range communication capabilities of our system. Our distance ranges from $1\hspace{1mm}mm$ to $10\hspace{1mm}km$ on the x-axis. Since distance itself is not the only parameter affecting $SER$, we tested the capabilities at different flow rates $[0.01, 0.1, 1]\hspace{1mm}m/s$ and at different $FNR$ values. Results indicate that to reach longer distances we need to have higher $FNR$ values. Fig. \ref{fig:fnr_analysis} also shows the effect of $FNR$ at different distances for different flow rates. 

\begin{figure}[b]
  \begin{subfigure}{0.45\linewidth}
    \centering
    \includegraphics[width=\linewidth]{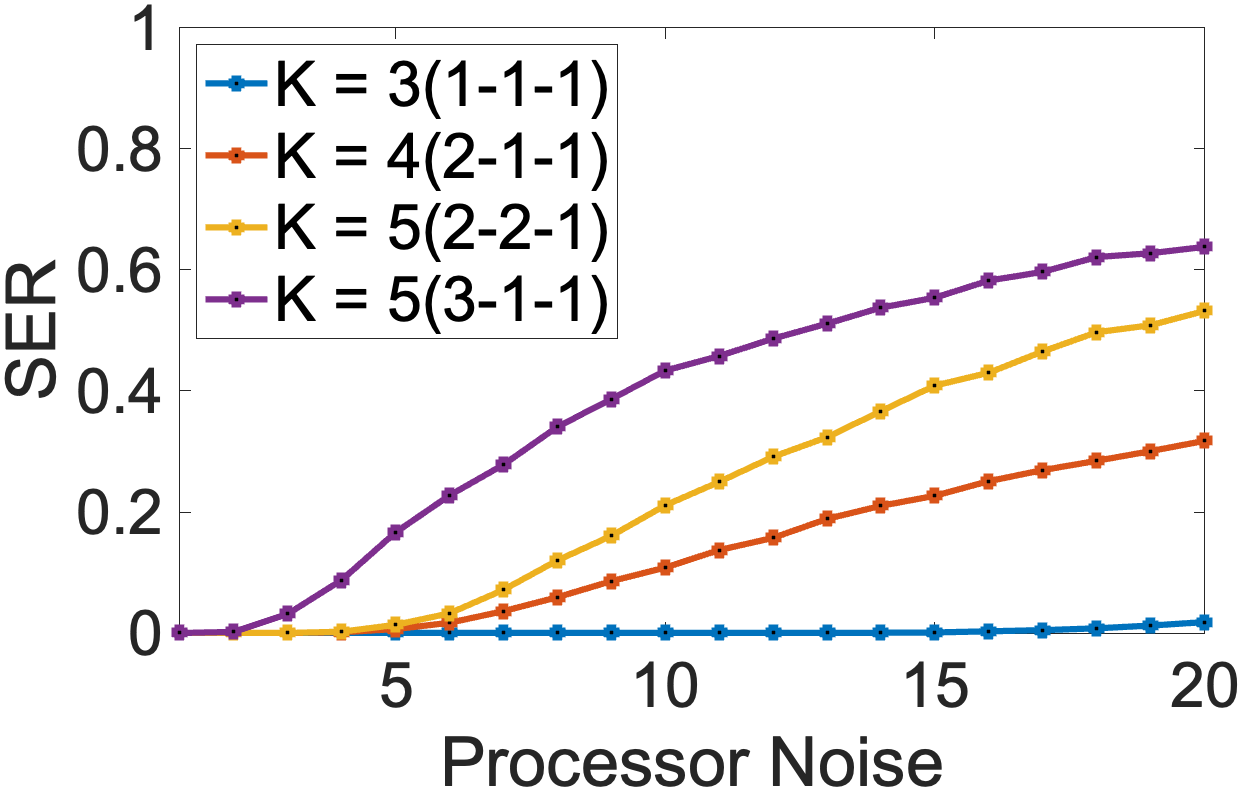}
    \caption{}
    \label{fig:processorNoiseAnalysis}
  \end{subfigure}
  \begin{subfigure}{0.44\linewidth}
    \centering
    \includegraphics[width=\linewidth]{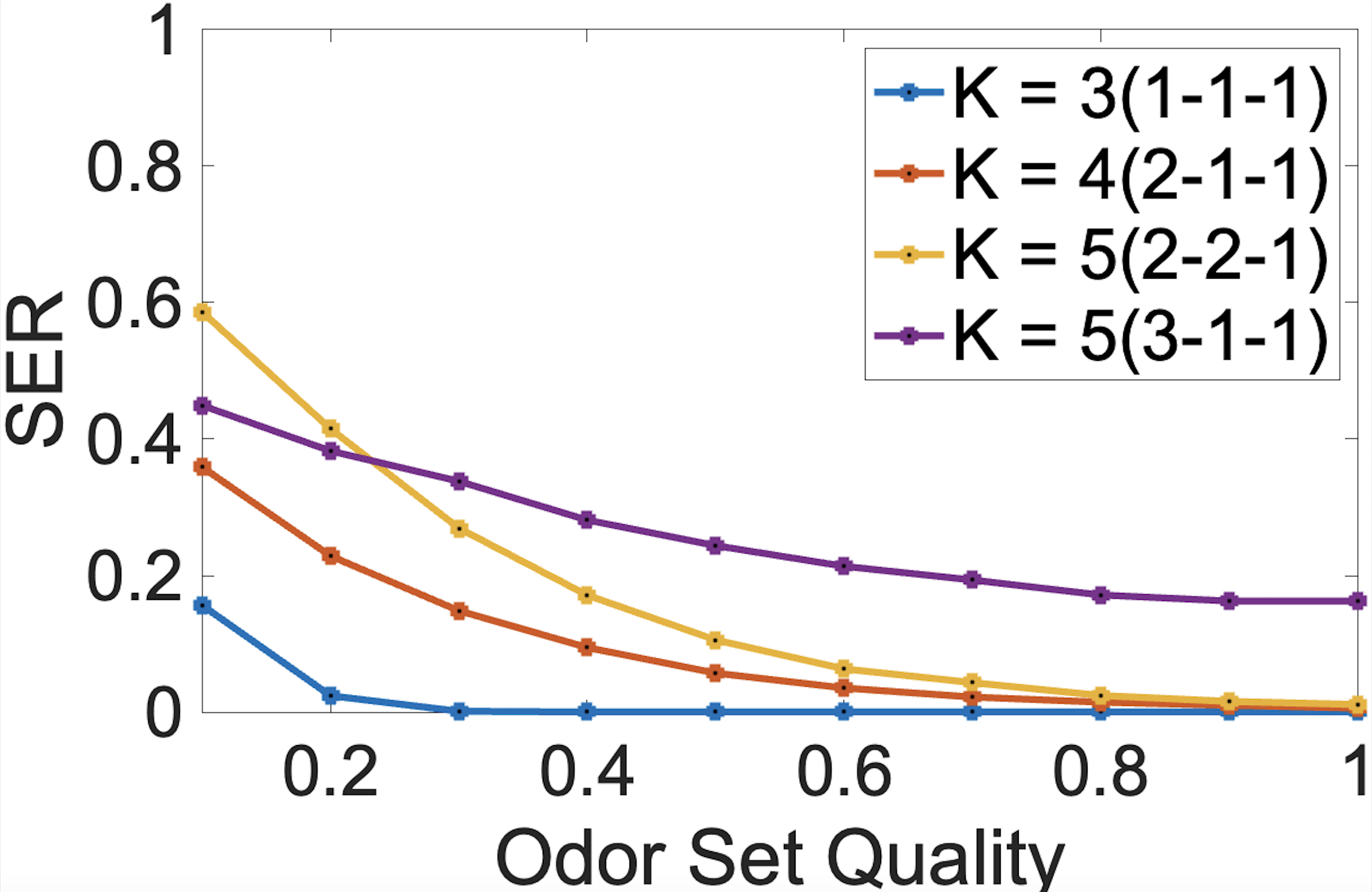}
    \caption{}
    \label{fig:qualityAnalysis}
  \end{subfigure}
  \caption{Type 2 $SER$ Analysis.}
  \label{fig:otherAnalyses}
\end{figure}

\subsubsection{Type 2 $SER$ Analysis}
For the analyses in this part, we tested our model with different combinations of the $K(n_p,n_i,n_e)$ where $K$ is the number of bits per symbol and $n_p, n_i$ and $n_e$ are the numbers of bits represented by the pleasantness, intensity, and edibility perceptual dimensions in a symbol. We created odor sets for the following combinations with the odor set qualities ranging from $0.1$ to $1$: $3(1,1,1)$, $4(2,1,1)$, $5(2,2,1)$, $5(3,1,1)$.

Then we firstly simulated the effect of $PN$ in Fig. \ref{fig:otherAnalyses}(a), while keeping the quality of odor sets $Q = 1$. General trend in the figure is increasing noise increases the $SER$. Moreover, as we increase the number of bits represented by a single perceptual dimension, OPSK becomes less prone to noise. Because decision regions in decoding phase becomes smaller. As a result, same amount of processor noise cause more decoding error and so more $SER$. 
The ambition behind increasing $K$ is to increase the number of bits for a single symbol so that we can reach to higher bits per second (bps) rates. If perceptual value vector of odors can be calculated more precisely with the help of future studies on this field, then we can even reach higher bps rates by further increasing the number of bits per symbol. This obviously requires $PN$ values in the range of $[0, 2]$. 

In figure \ref{fig:otherAnalyses}(b), we observe the effect of the quality of the odor set constituting odor bank. For the same $Q$ values we observe better $SER$ values for systems with lower $K$. The reason for that is again related to the area of the decision regions of the systems for different $Ks$. Since lower $Ks$ have larger decision regions, the quality of the odor set affects their performance with respect to ones with smaller decision regions.

\subsection{Symbol Rate Analysis}
\begin{figure}[t]
    \centering
    \includegraphics[width = 0.61\linewidth]{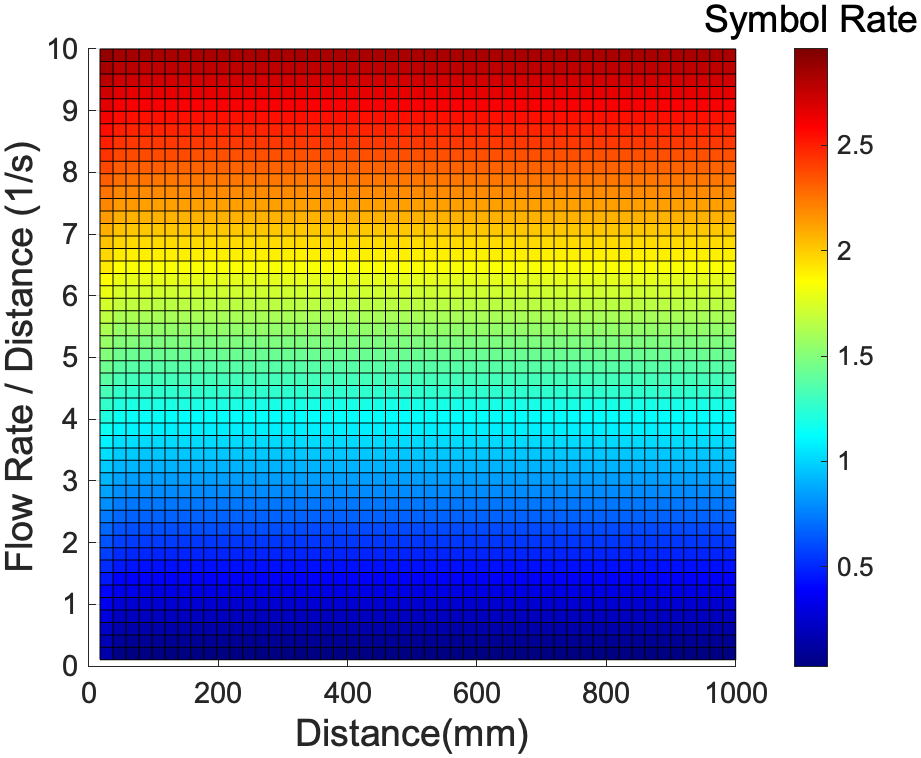}
    \caption{Symbol Rate Analysis.}
    \label{fig:symbolRateAnalysis}
\end{figure}

In this analysis, we investigated the symbol rate values for different distance and flow rate to distance ratios. Fig. \ref{fig:symbolRateAnalysis}, displays the symbol rate values at different colors mapped according to the color bar given in figure. The y-axis is chosen to be the flow rate to distance ratio so that the corresponding flow rate at a point in the figure can be calculated by multiplying the distance with the ratio. 

We observe an increase in the symbol rate as we increase flow rate to distance ratio for the same distance. This can be explained by increasing flow rate. However, from the previous analyses, we should also remember that we need to have better $FNR$ values at higher speeds. Hence, the system requires a trade-off: choosing either a higher data rate with lower quality or a lower data rate with higher quality.

On the other hand, if we keep the flow rate-to-distance ratio constant, increasing the distance results in similar $SER$ values because the flow rate increases in proportion to the distance. This behavior persists even when expanding the distance range to longer distances.

\begin{figure*}[b]
    \centering
    \begin{subfigure}{0.245\linewidth}
        \includegraphics[width=\linewidth]{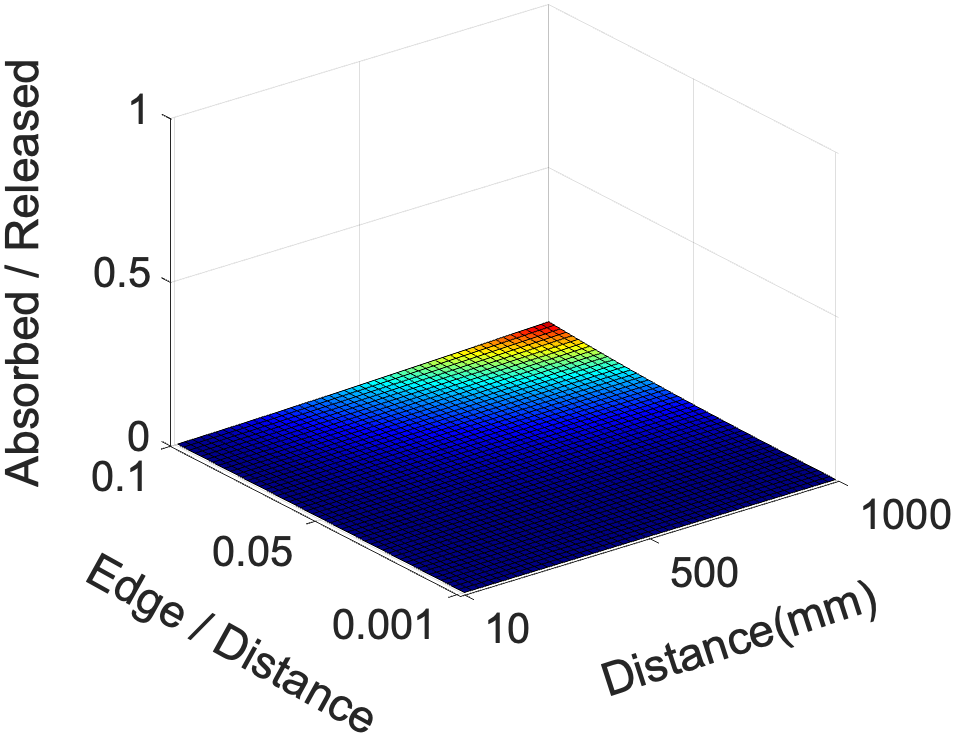}
        \caption{$u = 0.001 m/s$}
        \label{fig:massRatio_sub1}
    \end{subfigure}
    \begin{subfigure}{0.245\linewidth}
        \includegraphics[width=\linewidth]{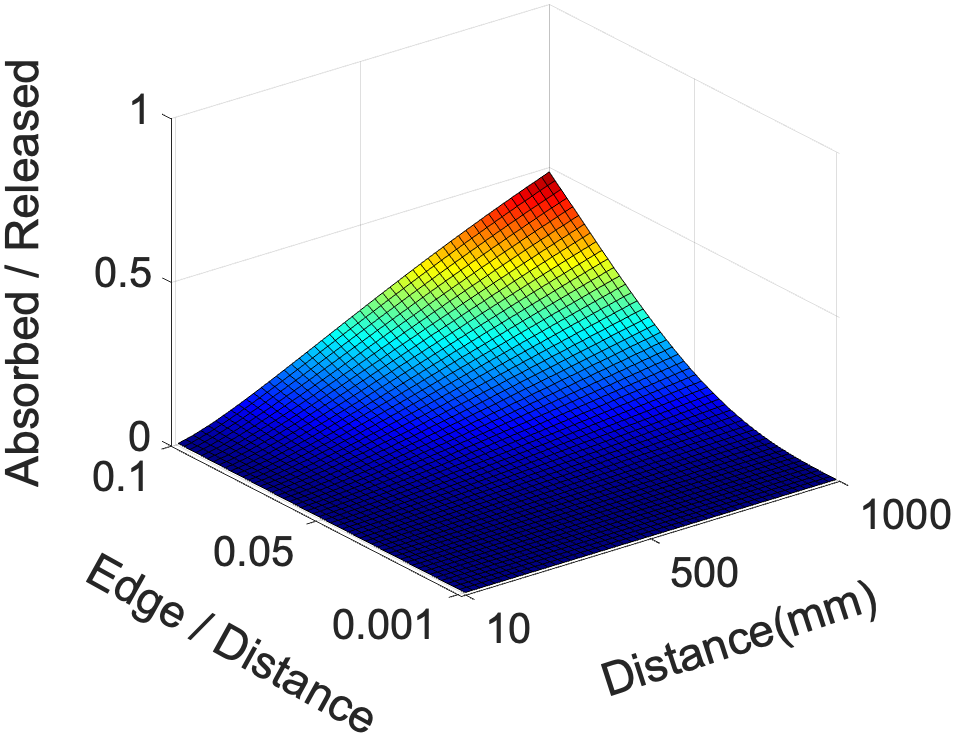}
        \caption{$u = 0.01 m/s$}
        \label{fig:massRatio_sub2}
    \end{subfigure}  
    \begin{subfigure}{0.245\linewidth}
        \includegraphics[width=\linewidth]{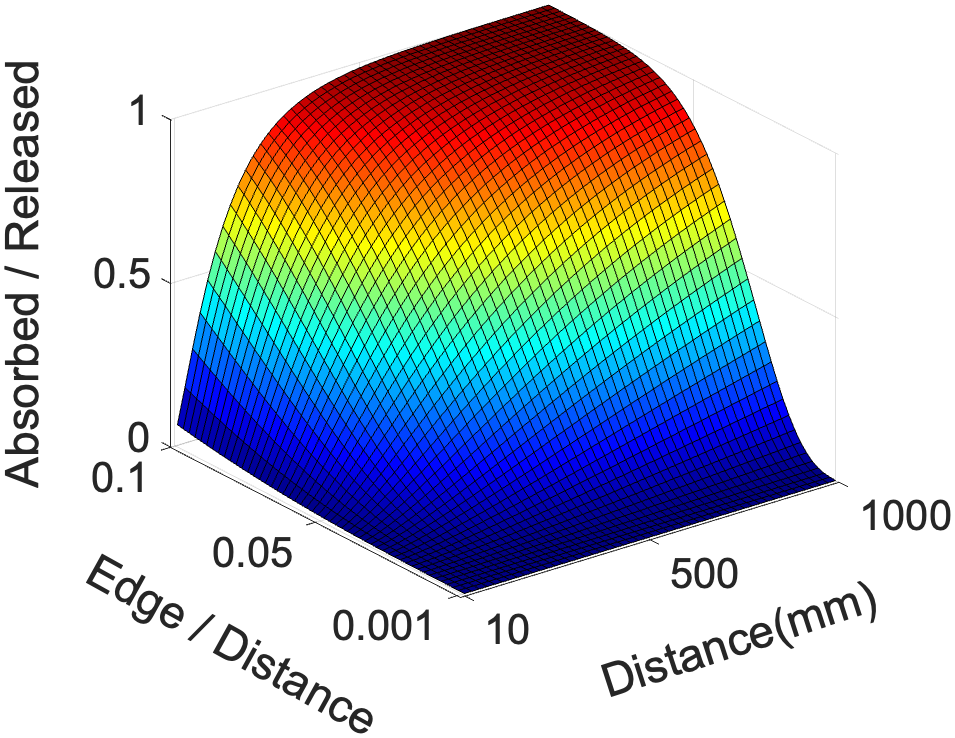}
        \caption{$u = 0.1 m/s$}
        \label{fig:massRatio_sub3}
    \end{subfigure}
    \begin{subfigure}{0.245\linewidth}
        \includegraphics[width=\linewidth]{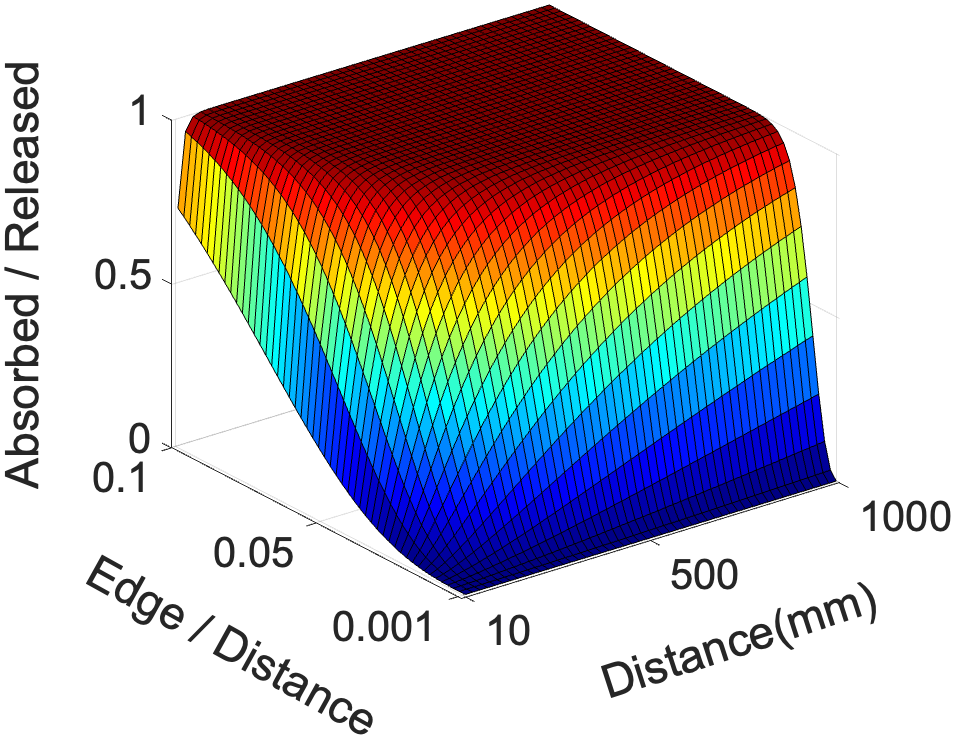}
        \caption{$u = 1 m/s$}
        \label{fig:massRatio_sub4}
    \end{subfigure}
    \caption{Mass Ratio Analysis for different flow rates $\boldsymbol{\vec{u}}$ where $\boldsymbol{\vec{u}} = (u, u, u)$.}
    \label{fig:massRatioAnalysis}
\end{figure*}

\subsection{Mass Ratio Analysis}\label{Section:Mass Ratio Analysis}
This analysis is especially useful for the future experimental studies on this field. For the demodulation operation explained in \ref{Receiver:Demodulation}, the processor unit would need a minimum mass to calculate the perceptual value vector of the odor with the greatest mass. This analysis addresses how much mass to release from the transmitter according to the proportion of the released mass that is being absorbed. 

We defined the performance parameter mass ratio as coefficient of $M_{n+1-k}$  when $k = 1$ in (\ref{eq:mn}). This ratio indicate the proportion of the released mass that is being absorbed. For this analysis, we observed the effect of multiple system parameters like distance between transmitter and receiver, the edge length of the cube-shaped receiver, and flow rate in the channel, at the same time. 

Fig. \ref{fig:massRatioAnalysis}, displays the analysis results at each subplot for different flow rate values. Each subplot has distance ranges from $10\hspace{1mm}mm$ to $1\hspace{1mm}m$ on the x-axis, edge length to distance ratio ranges from $0.001$ to $0.1$ on the y-axis and mass ratio on the z-axis. Mass ratios on the z-axis are the expected values. Because of the channel noise, deviations from these ratios occur which lead to symbol errors. 

For each different flow rate values at Fig. \ref{fig:massRatioAnalysis}, we observe that increasing edge length to distance ratio increases the mass ratio when distance kept constant. This is because the receiver absorbs molecules from bigger volume. On the other hand, we observe that increasing distance for the same edge length to distance ratio increases the mass ratio. Because this also implies an increase on receiver’s edge length and so the volume. When we evaluate plots all together, we observe that increasing flow rate gradually from $0.001\hspace{1mm}m/s$ to $1\hspace{1mm}m/s$ increases the mass ratio. This can be explained as, increasing flow rate reduces the effect of diffusion and dispersed molecules stay in a more limited area.

\subsection{Operation Time Extension Analysis for Adaptive Symbol Transmission}

In this analysis, we explored how adaptive symbol transmission affects the extension of operation time. For this we created communication systems for the following combinations: $3(1,1,1)$, $4(2,1,1)$, $5(2,2,1)$, $5(3,1,1)$. We assessed the effects of using adaptive transmitters and receivers across each combination by examining four distinct frequency distributions of the symbols set to be transmitted. The first distribution presents a scenario where all symbols have an equal probability of occurring within the sequence, while the remaining distributions are randomly generated, featuring symbols that do not occur with equal probability.

\begin{figure*}[t]
    \centering
    \begin{subfigure}{0.245\linewidth}
        \includegraphics[width=\linewidth]{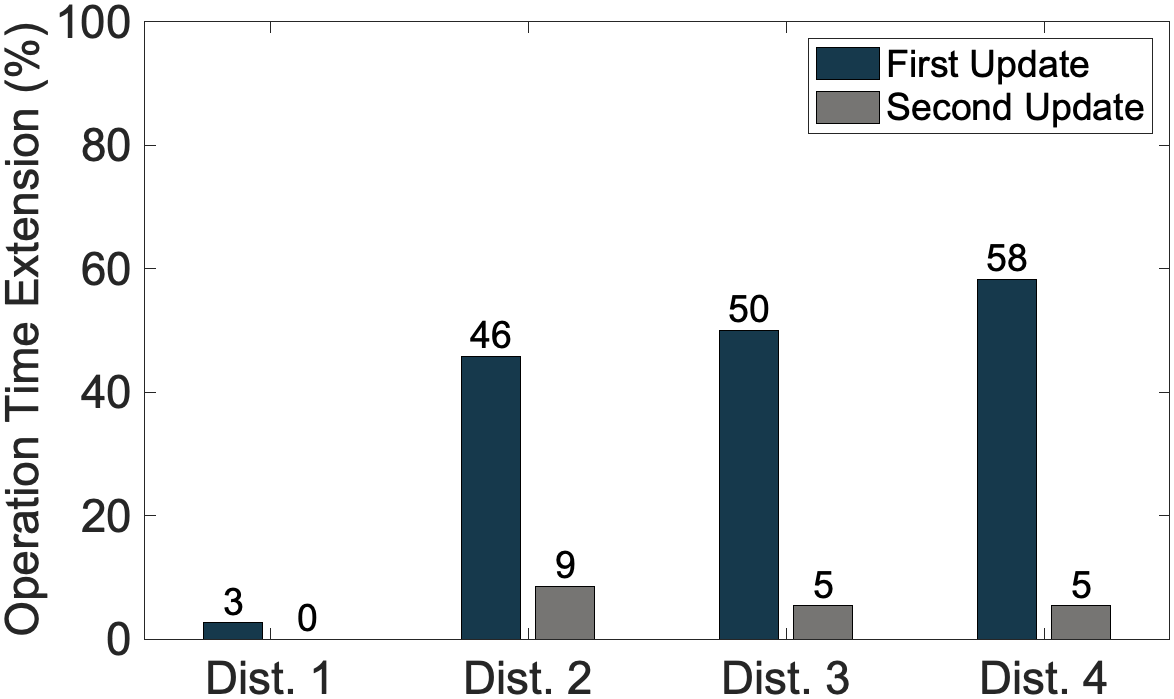}
        \caption{Starting State : $3(1,1,1)$}
    \end{subfigure}
    \begin{subfigure}{0.245\linewidth}
        \includegraphics[width=\linewidth]{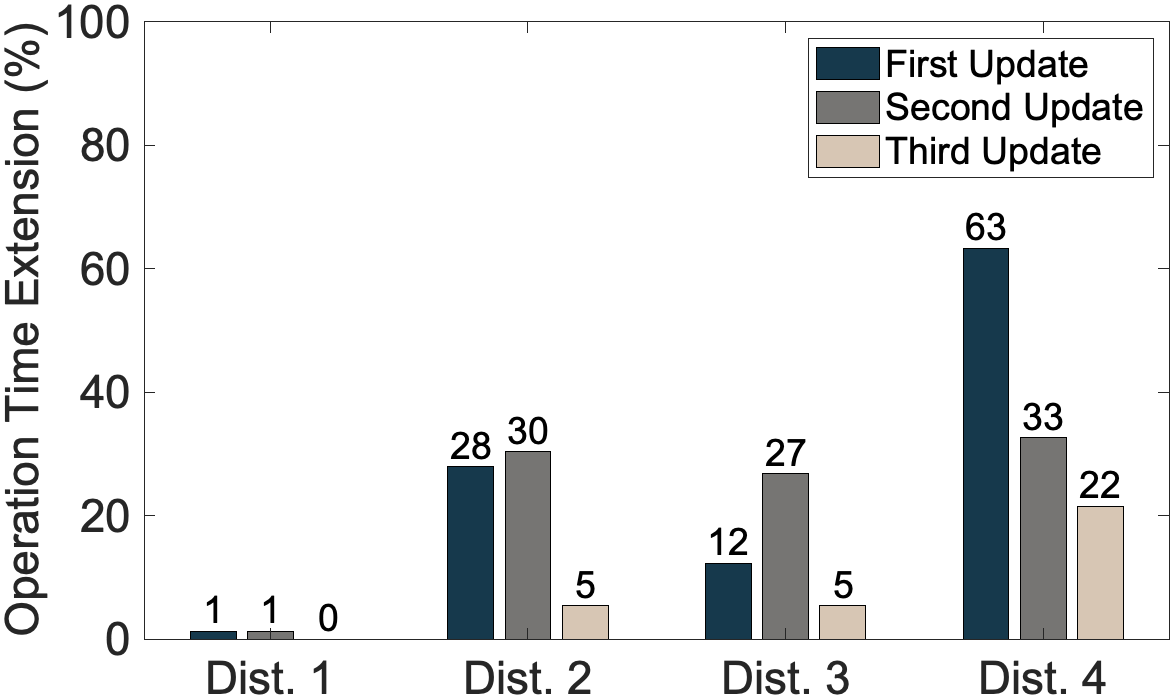}
        \caption{Starting State : $4(2,1,1)$}
    \end{subfigure}  
    \begin{subfigure}{0.245\linewidth}
        \includegraphics[width=\linewidth]{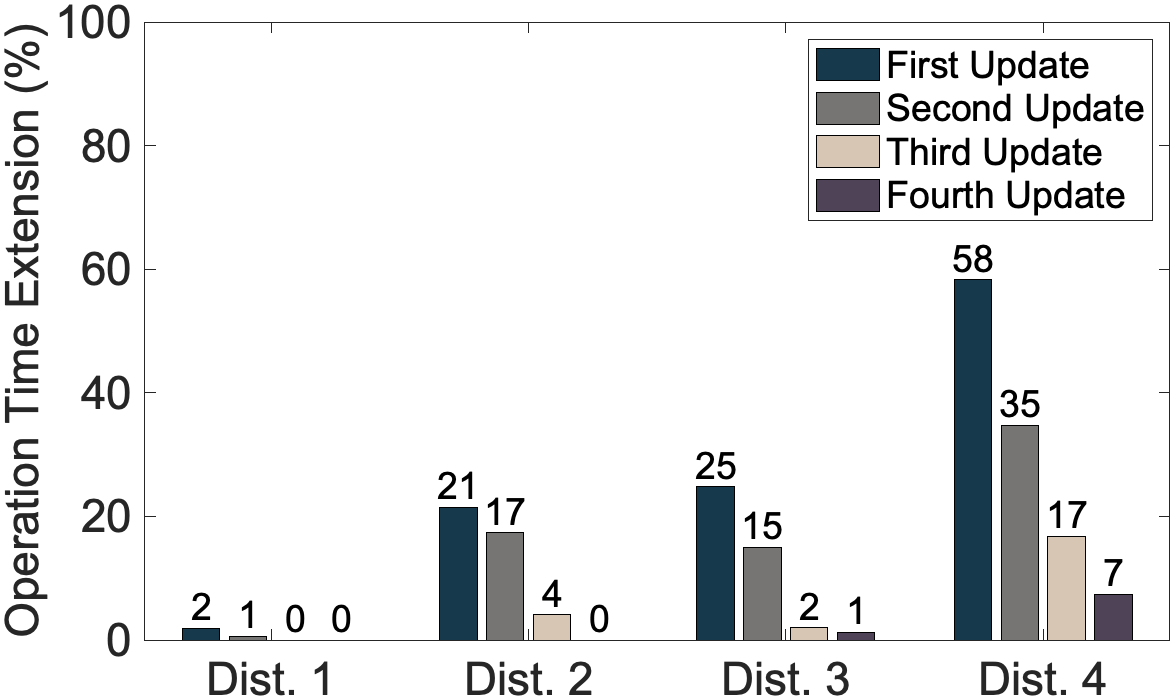}
        \caption{Starting State : $5(2,2,1)$}
    \end{subfigure}
    \begin{subfigure}{0.245\linewidth}
        \includegraphics[width=\linewidth]{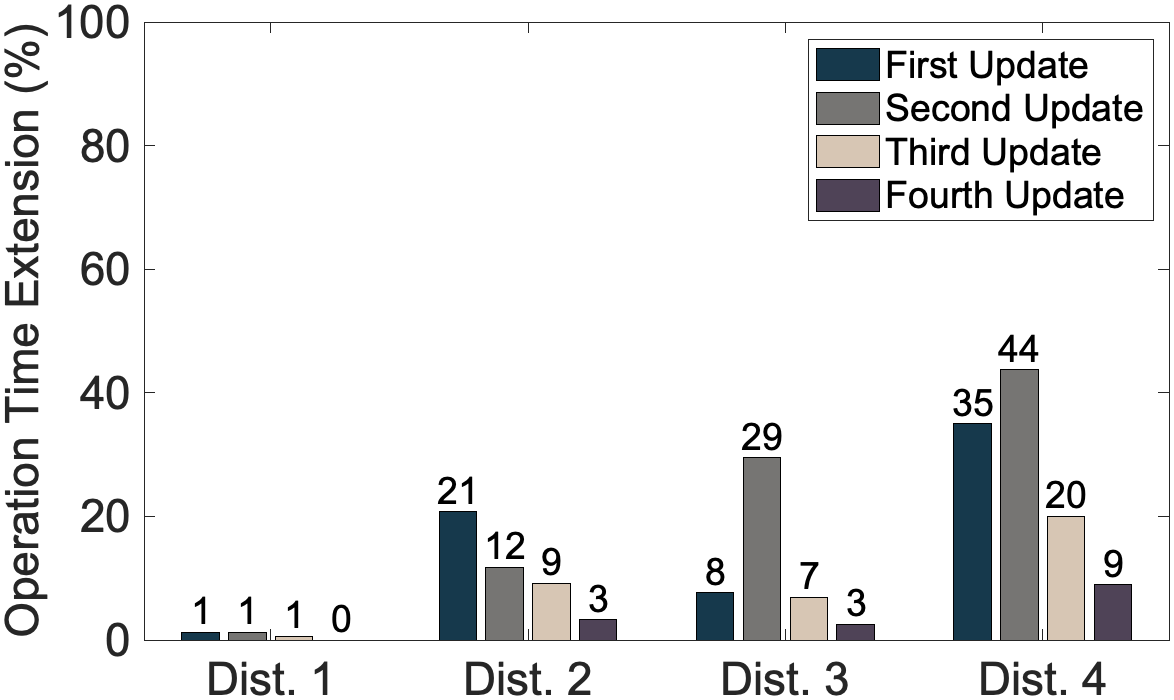}
        \caption{Starting State : $5(3,1,1)$}
    \end{subfigure}
    \caption{Operation Time Extension Analysis}
    \label{fig:adaptiveSymbolTransmissionAnalysis}
\end{figure*}

Fig. \ref{fig:adaptiveSymbolTransmissionAnalysis} displays the results of operation time extensions for each combination and distribution, depicted through bar graphs. The values represented in the bars are derived from a straightforward procedure detailed in Algorithm \ref{algorithm}. The system's runtime is calculated until it reaches its first update state. Then, the optimal possible extension time is estimated according to the most recently transmitted N symbols. Following that, the ratio of the potential extension time to the initial runtime is expressed as a percentage. After the first update, this procedure is iterated. The system is updated whenever it became incapable of transmitting N more symbols, based on the statistics of the most recently transmitted N symbols.

To maximize the extension of operation time without restrictions, we pursued updates as long as any feasible improvement was achievable. The updating process is stopped when the system configuration reached 1 bit per symbol, at which point further updates became impossible because each subsequent update would decrease the number of bits per symbol.

Fig. \ref{fig:adaptiveSymbolTransmissionAnalysis} reveals that system updates in the first distribution for each combination do not enhance operation time extension. This outcome arises because symbols that occur with equal probability in the sequence cause odors from different classes to deplete simultaneously. Conversely, for the other distributions across all combinations, significant improvements in the system's operation time are observed. The reason behind this is that odors from certain classes deplete more quickly, and merging odor classes through updates allows the system to operate for an extended period.

\section{Conclusion}

In this paper, we introduced a novel modulation technique called OPSK for odor-based communication. We constructed a system model and described the parts of system from transmitter to receiver. With the analyses done on the constructed system, we tested the capabilities of our system under many different conditions. Analyses have shown that at distances on the scale of millimeters, the system operates successfully even in the presence of considerable noise. However, when communicating over longer distances, such as kilometers, while aiming for low $SER$ values, the noise in the channel becomes a significant concern. It is also observed that at higher flow rates, the system becomes more susceptible to noise.

Another observation from our analyses was the correlation between the increase in symbol rate and the rise in flow rate. Alternatively, data rate can also be increased by representing more bits with each symbol. However, it was noted that representing a higher number of bits per dimension necessitates greater accuracy in the demodulation process. Therefore, to further enhance the data rate, conducting more research into precisely measuring the perceptual values of odor molecules would be highly beneficial. In addition to this, increasing the number of perceptual dimensions offers another way to increase the data rate, although this area requires further research for effective implementation, too.

When developing our point-to-point communication system, we also considered its potential suitability for creating multi-node systems. In such setups, each node could be uniquely addressed using a specific perceptual dimension of odors. This way, each node would only interact with odor molecules that match its address. This approach could pave the way for highly selective and efficient network communications in MC context.

\bibliographystyle{IEEEtran}
\bibliography{references.bib}

\begin{thebibliography}{10}
\providecommand{\url}[1]{#1}
\csname url@rmstyle\endcsname
\providecommand{\newblock}{\relax}
\providecommand{\bibinfo}[2]{#2}
\providecommand\BIBentrySTDinterwordspacing{\spaceskip=0pt\relax}
\providecommand\BIBentryALTinterwordstretchfactor{4}
\providecommand\BIBentryALTinterwordspacing{\spaceskip=\fontdimen2\font plus
\BIBentryALTinterwordstretchfactor\fontdimen3\font minus \fontdimen4\font\relax}
\providecommand\BIBforeignlanguage[2]{{%
\expandafter\ifx\csname l@#1\endcsname\relax
\typeout{** WARNING: IEEEtran.bst: No hyphenation pattern has been}%
\typeout{** loaded for the language `#1'. Using the pattern for}%
\typeout{** the default language instead.}%
\else
\language=\csname l@#1\endcsname
\fi
#2}}

\bibitem{mcsurvey}
N.~Farsad, H.~B. Yilmaz, A.~Eckford, C.-B. Chae, and W.~Guo, ``A comprehensive survey of recent advancements in molecular communication,'' \emph{IEEE Communications Surveys \& Tutorials}, vol.~18, no.~3, pp. 1887--1919, 2016.

\bibitem{akanmcsurvey}
O.~B. Akan, H.~Ramezani, T.~Khan, N.~A. Abbasi, and M.~Kuscu, ``Fundamentals of molecular information and communication science,'' \emph{Proceedings of the IEEE}, vol. 105, no.~2, pp. 306--318, 2017.

\bibitem{modsurvey}
M.~S. Kuran, H.~B. Yilmaz, I.~Demirkol, N.~Farsad, and A.~Goldsmith, ``A survey on modulation techniques in molecular communication via diffusion,'' \emph{IEEE Communications Surveys \& Tutorials}, vol.~23, no.~1, pp. 7--28, 2021.

\bibitem{plantinsect}
L.~Conchou, P.~Lucas, C.~Meslin, M.~Proffit, M.~Staudt, and M.~Renou, ``Insect odorscapes: From plant volatiles to natural olfactory scenes,'' \emph{Frontiers in Physiology}, vol.~10, 2019.

\bibitem{insect}
M.~Renou and S.~Anton, ``Insect olfactory communication in a complex and changing world,'' \emph{Current Opinion in Insect Science}, vol.~42, pp. 1--7, 2020.

\bibitem{human}
M.~Russell, ``Human olfactory communication,'' \emph{Nature}, vol. 260, pp. 520--522, 1976.

\bibitem{longrangeMC}
L.~Gine and I.~Akyildiz, ``Molecular communication options for long range nanonetworks,'' \emph{Computer Networks}, vol.~53, pp. 2753--2766, 2009.

\bibitem{mammals}
J.~F. Eisenberg and D.~G. Kleiman, ``Olfactory communication in mammals,'' \emph{Annual Review of Ecology and Systematics}, vol.~3, no.~1, pp. 1--32, 1972.

\bibitem{odorcsk}
Y.~Ariyakul and S.~Wisayataksin, ``Data modulation technique using concentration of odor molecules,'' pp. 1--4, 2018.

\bibitem{cskmsk}
M.~S. Kuran, H.~B. Yilmaz, T.~Tugcu, and I.~F. Akyildiz, ``Modulation techniques for communication via diffusion in nanonetworks,'' pp. 1--5, 2011.

\bibitem{shortmod1}
N.-R. Kim and C.-B. Chae, ``Novel modulation techniques using isomers as messenger molecules for nano communication networks via diffusion,'' \emph{IEEE Journal on Selected Areas in Communications}, vol.~31, no.~12, pp. 847--856, 2013.

\bibitem{shortmod2}
S.~Pudasaini, S.~Shin, and K.~Kwak, ``Run-length aware hybrid modulation scheme for diffusion-based molecular communication,'' \emph{14th International Symposium on Communications and Information Technologies, ISCIT 2014}, 2014.

\bibitem{shortmod3}
------, ``Robust modulation technique for diffusion-based molecular communications in nanonetworks,'' vol.~X, pp. 1--4, 2015.

\bibitem{shortmod4}
M.~C. Gursoy, E.~Basar, A.~E. Pusane, and T.~Tugcu, ``Index modulation for molecular communication via diffusion systems,'' \emph{IEEE Transactions on Communications}, vol.~67, no.~5, pp. 3337--3350, 2019.

\bibitem{shortmod5}
------, ``Pulse position-based spatial modulation for molecular communications,'' \emph{IEEE Communications Letters}, vol.~23, no.~4, pp. 596--599, 2019.

\bibitem{shortmod6}
Y.~Tang, M.~Wen, X.~Chen, Y.~Huang, and L.-L. Yang, ``Molecular type permutation shift keying for molecular communication,'' \emph{IEEE Transactions on Molecular, Biological and Multi-Scale Communications}, vol.~6, no.~2, pp. 160--164, 2020.

\bibitem{shortmod7}
D.~Aktas and O.~B. Akan, ``Weight shift keying (wsk) with practical mechanical receivers for molecular communications in internet of everything,'' \emph{IEEE Journal on Selected Areas in Communications}, vol.~40, no.~11, pp. 3285--3294, 2022.

\bibitem{kuscutransmittersurvey}
M.~Kuscu, E.~Dinc, B.~A. Bilgin, H.~Ramezani, and O.~B. Akan, ``Transmitter and receiver architectures for molecular communications: A survey on physical design with modulation, coding, and detection techniques,'' \emph{Proceedings of the IEEE}, vol. 107, no.~7, pp. 1302--1341, 2019.

\bibitem{olfactionhistory}
C.~M. Philpott, A.~Bennett, and G.~E. Murty, ``A brief history of olfaction and olfactometry,'' \emph{The Journal of Laryngology \& Otology}, vol. 122, no.~7, p. 657–662, 2008.

\bibitem{dyson}
R.~H. Wright, ``Odour and chemical constitution,'' \emph{Nature}, vol. 173, p. 831, 1954.

\bibitem{stereochemical}
J.~Amoore, ``Stereochemical theory of olfaction,'' \emph{Nature}, vol. 198, p. 271–272, 1963.

\bibitem{dravnieks}
A.~Dravnieks, ``Physicochemical basis of olfaction,'' \emph{Annals of the New York Academy of Sciences}, vol. 116, p. 429–439, 1964.

\bibitem{pleasantnesslinear}
M.~Zarzo, ``Hedonic judgments of chemical compounds are correlated with molecular size,'' \emph{Sensors (Basel, Switzerland)}, vol.~11, pp. 3667--86, 2011.

\bibitem{dravnieksatlas}
A.~Dravnieks, ``Atlas of odor character profiles,'' \emph{American Society for Testing and Materials}, 1985.

\bibitem{harper}
R.~Harper, D.~G. Land, and N.~M. Griffiths, ``Odour qualities: A glossary of usage,'' \emph{British Journal of Psychology}, vol.~59, no.~3, pp. 231--252, 1968.

\bibitem{moncrieff}
R.~W. Moncrieff, \emph{Odour Preferences}.\hskip 1em plus 0.5em minus 0.4em\relax Leonard Hill, 1966.

\bibitem{wrightdata}
R.~H. Wright and K.~M. Michels, ``Evaluation of far infrared relations to odor by a standards similarity method,'' \emph{Annals of the New York Academy of Sciences}, vol. 116, pp. 535--551, 1964.

\bibitem{amoore1967correlations}
J.~E. Amoore and D.~Venstrom, ``Correlations between stereochemical assessments and organoleptic analysis of odorous compounds,'' \emph{Olfaction and Taste II}, pp. 3--17, 1967.

\bibitem{pleasant}
R.~Khan, C.-H. Luk, A.~Flinker, A.~Aggarwal, H.~Lapid, R.~Haddad, and N.~Sobel, ``Predicting odor pleasantness from odorant structure: Pleasantness as a reflection of the physical world,'' \emph{The Journal of neuroscience : The Official Journal of the Society for Neuroscience}, vol.~27, pp. 10\,015--23, 2007.

\bibitem{zarzopleasantnessedibility}
M.~Zarzo, ``Psychologic dimensions in the perception of everyday odors: Pleasantness and edibility,'' \emph{Journal of Sensory Studies}, vol.~23, pp. 354--376, 2008.

\bibitem{intensitypercept}
C.~Bontempi, P.~Corbelin, G.~Brand, and L.~Jacquot, ``Ortho- and retronasal stimulations with specific food odours: Hedonic and familiarity ratings are related to chemosensory pleasure scale scores,'' \emph{Flavour and Fragrance Journal}, vol.~38, no.~4, pp. 243--252, 2023.

\bibitem{intensityperceptual}
A.~Bierling, I.~Croy, T.~Hummel, G.~Cuniberti, and A.~Croy, ``Olfactory perception in relation to the physicochemical odor space,'' \emph{Brain Sciences}, vol.~11, p. 563, 2021.

\bibitem{intensityperceptual2}
A.~Keller and L.~Vosshall, ``Olfactory perception of chemically diverse molecules,'' \emph{BMC Neuroscience}, vol.~17, p.~55, 2016.

\bibitem{intensity}
S.~Schiffman, R.~Gutierrez-Osuna, and H.~Nagle, Jr, ``Measuring odor intensity with e-noses and other sensor types,'' \emph{Proceedings of the 9th International Symposium on Olfaction and Electronic Nose}, 2002.

\bibitem{enosepleasantness}
R.~Haddad, A.~Medhanie, Y.~Roth, D.~Harel, and N.~Sobel, ``Predicting odor pleasantness with an electronic nose,'' \emph{PLoS Computational Biology}, vol.~6, p. e1000740, 2010.

\bibitem{olfactionbase}
\BIBentryALTinterwordspacing
A.~Sharma, B.~K. Saha, R.~Kumar, and P.~K. Varadwaj, ``Olfactionbase: a repository to explore odors, odorants, olfactory receptors and odorant–receptor interactions,'' \emph{Nucleic Acids Research}, vol.~50, no.~D1, p. D678–D686, 2022. [Online]. Available: \url{https://doi.org/10.1093/nar/gkab763}
\BIBentrySTDinterwordspacing

\bibitem{tuncparam}
D.~T. McGuiness, S.~Giannoukos, A.~Marshall, and S.~Taylor, ``Parameter analysis in macro-scale molecular communications using advection-diffusion,'' \emph{IEEE Access}, vol.~6, pp. 46\,706--46\,717, 2018.

\bibitem{advectioneq}
\BIBentryALTinterwordspacing
C.~E. Baukal, V.~Y. Gershtein, and X.~Li, \emph{Computational Fluid Dynamics in Industrial Combustion}, 2001. [Online]. Available: \url{https://api.semanticscholar.org/CorpusID:106632286}
\BIBentrySTDinterwordspacing

\bibitem{gaussian}
R.~Roy, P.~Chattopadhyay, B.~Tudu, N.~Bhattacharyya, and R.~Bandyopadhyay, ``Artificial flavor perception of black tea using fusion of electronic nose and tongue response: A bayesian statistical approach,'' \emph{Journal of Food Engineering}, vol. 142, p. 87–93, 2014.

\end{thebibliography}

\newpage
\vspace{-3cm}
\begin{IEEEbiography}
    [{\includegraphics[width=1in, height=1.25in, clip, keepaspectratio]{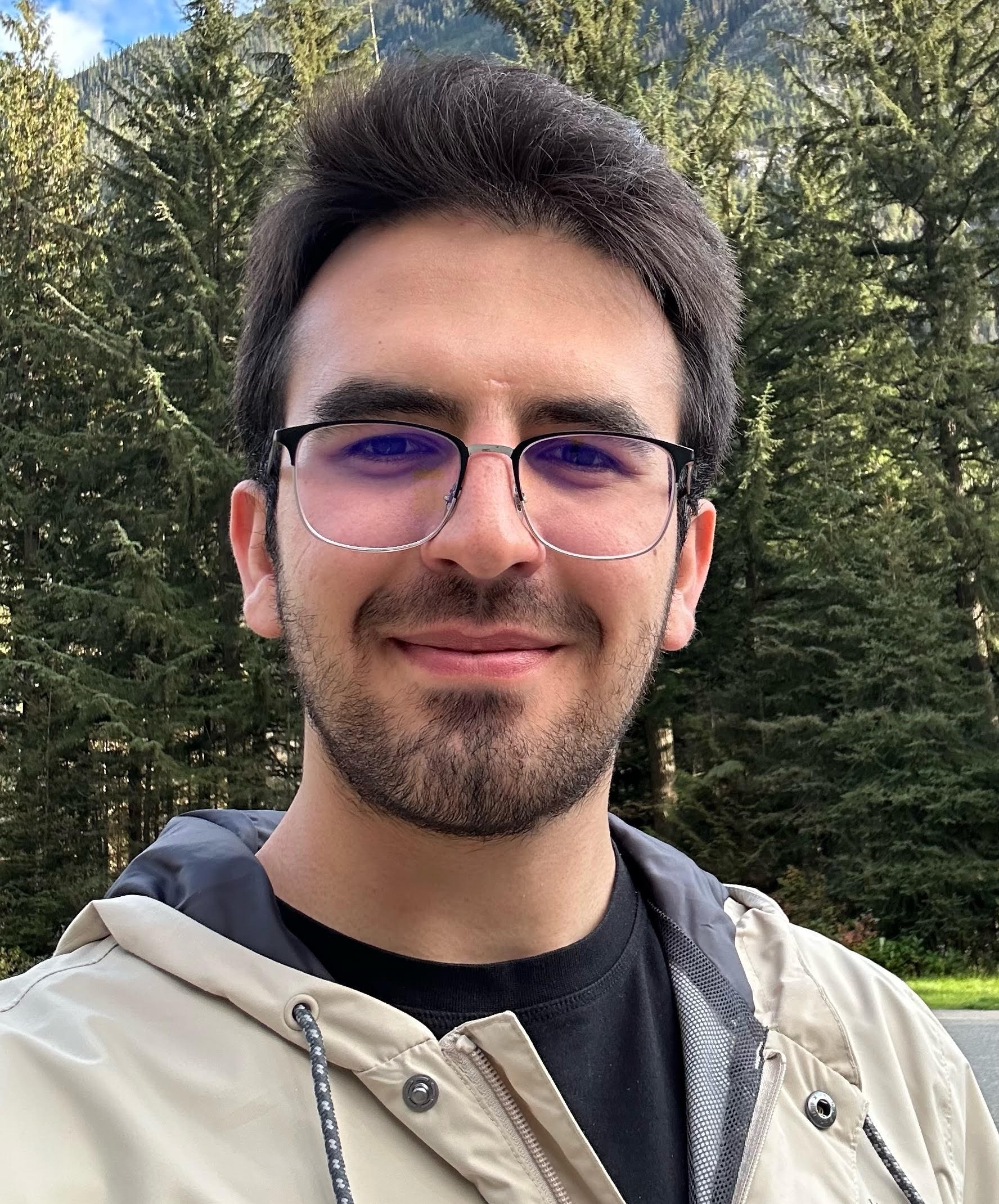}}]{Fatih Efe Bilgen} completed his high school education at Kocaeli Enka Science and Technology High School with a silver medal in National Physics Olympiads. He is currently a 4th year student in Electrical and Electronics Engineering with a double major in Mathematics at Koç University, Istanbul, Turkey. He is a research assistant at the Center for neXt-generation Communications (CXC) under the supervision of Prof. Akan.
\end{IEEEbiography}
\vspace{-11cm}
\begin{IEEEbiography}
    [{\includegraphics[width=1in, height=1.25in, clip, keepaspectratio]{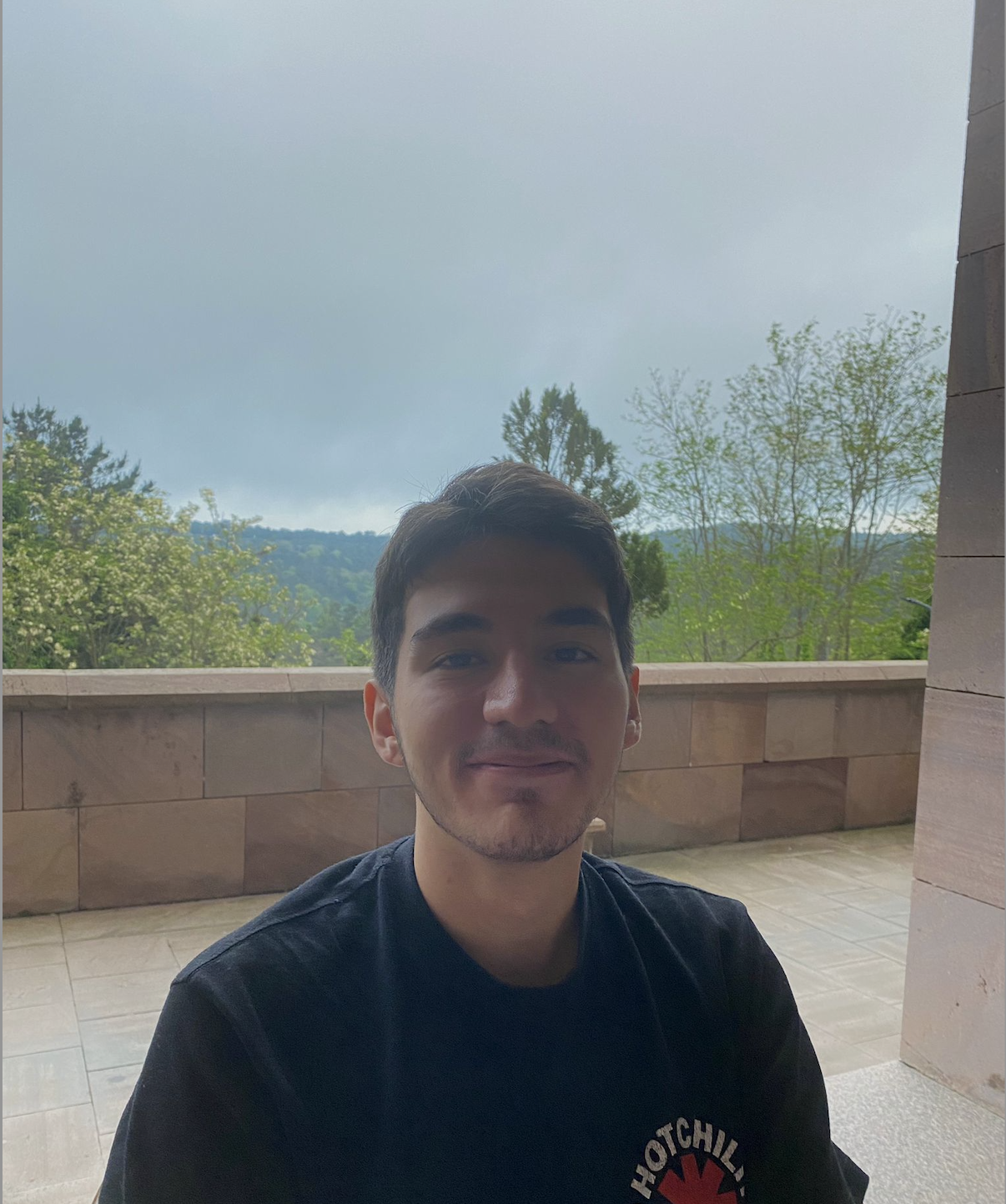}}]{Ahmet Burak Kilic} completed his high school education Bilfen Kayseri High School, Kayseri, Turkey. He is currently a senior student in Electrical and Electronics Engineering with a double major in Mathematics at Koç University, Istanbul, Turkey. He is a research assistant at the Center for neXt-generation Communications (CXC) under the supervision of Prof. Akan.
\end{IEEEbiography}
\vspace{-11cm}
\begin{IEEEbiography}
[{\includegraphics[width=1in,height=1.25in,clip,keepaspectratio]{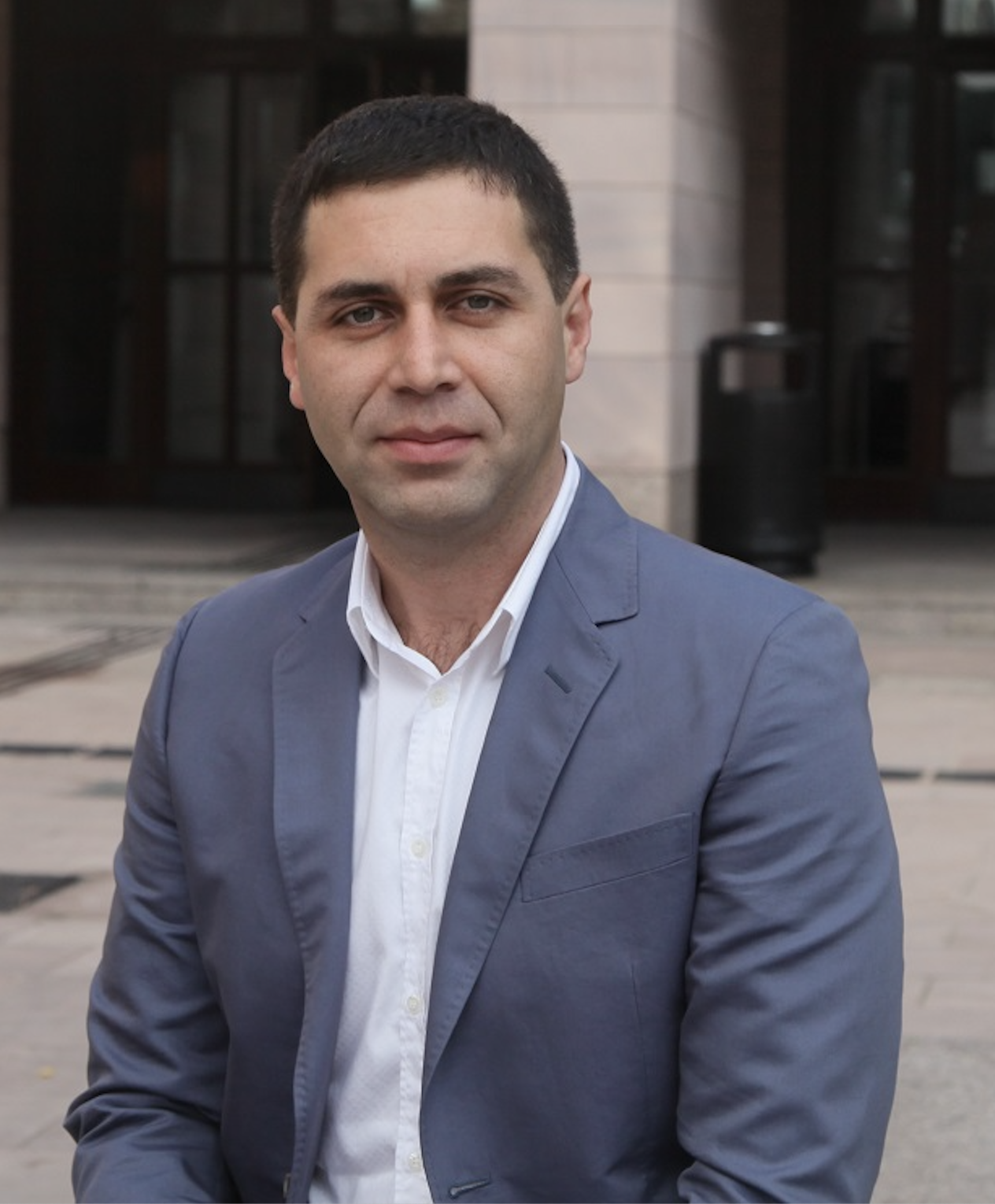}}]{Ozgur B. Akan (Fellow, IEEE)}
received the PhD from the School of Electrical and Computer Engineering Georgia Institute of Technology Atlanta, in 2004. He is currently the Head of Internet of Everything (IoE) Group, with the Department of Engineering, University of Cambridge, UK and the Director of Centre for neXt-generation Communications (CXC), Koç University, Turkey. His research interests include wireless, nano, and molecular communications and Internet of Everything.
\end{IEEEbiography}
\end{document}